\documentclass[twocolumn]{aastex631}

\usepackage{newtxtext,newtxmath}
\usepackage{verbatim}
\usepackage{color}
\usepackage{makecell}
\usepackage{enumerate}

\begin{document}

\title{Quasi-periodic oscillations and reflection feature evolution in 4U 1630-47 observed with Insight-HXMT}

\correspondingauthor{Wei Wang}
\email{wangwei2017@whu.edu.cn}

\author[0000-0001-5514-1167]{Jiashi Chen}
\affiliation{Department of Astronomy, School of Physics and Technology,
Wuhan University, Wuhan 430072, China}

\author[0000-0003-3901-8403]{Wei Wang}
\affiliation{Department of Astronomy, School of Physics and Technology, Wuhan University, Wuhan 430072, China}

\begin{abstract}
The Galactic black hole (BH) X-ray binary 4U 1630-47 went into a new outburst in 2021 after $\sim$ 600 days from its 2020 outburst. We perform a detailed analysis of quasi-periodic oscillations and spectral evolutions during its 2021 outburst based on \textit{Insight}-HXMT observations. The main science aims to study the reflection features evolution of this accreting black hole using the observations of detecting quasi-periodic oscillations (QPOs) and quasi-regular modulations (QRMs). The QPOs frequencies evolve from $\sim 1.6 - 3.6$ Hz, and QRMs have low frequencies around 0.05 - 0.07 Hz. The reflection fraction varies during the outburst and has a positive correlation with the hardness ratio when QPOs are detected. The centroid frequency of QPOs is anti-correlated to the reflection fraction. This is consistent with the prediction of precessing inner flow model and provides evidence for a geometrical origin of QPOs. The centroid frequency of QRMs also shows an anti-correlation to the reflection fraction, but the hardness ratio shows no relation to the reflection fraction during the period. We suggest that QRMs may have a different origin from QPOs and be caused by instabilities in the corona.

\end{abstract}


\section{Introduction}

Most black hole X-ray binaries are transient sources, characterized by alternating periods of dormancy marked by low flux or quiescent states, and outbursts featuring high flux or active states over their evolutionary trajectory. These transient systems primarily inhabit a quiescent phase with emissions remaining undetectable until the onset of outbursts. During outbursts, the X-ray emission experiences a significant surge, generally raising several orders of magnitude and displaying distinct spectral states (see \citealt{2006ARA&A..44...49R} for a review). During an outburst, black hole X-ray binaries typically cycle through several spectral states that can be traced using a hardness-intensity diagram (HID). In the HID, on the abscissa is the harness, which is defined as the ratio of the counts in a harder X-ray band to that in a softer X-ray band; on the ordinate is the total count rate over a broad energy band, which gives a proxy for luminosity and accretion rate (see, e.g., \citealt{2001ApJS..132..377H,2005A&A...440..207B,2005ApJ...624..295H}). The general evolution of an outburst in HID is a "q" shaped diagram that travels counterclockwise from the bottom right (see Figure 3 in \citealt{2006ARA&A..44...49R}). The right vertical branch corresponds to the Low/Hard State (LHS) that characterized by large variability and a hard spectrum, which is observed at the start and the end of an outburst only. The left vertical branch corresponds to the High/Soft State (HSS), whose spectrum is soft, and the variability is low. The central part of the diagram identifies two intermediate states (IMS), the Hard Intermediate State (HIMS) and Soft Intermediate State (SIMS). During the IMS, the X-ray spectrum exhibits characteristics of both the LHS and the HSS. Although for different sources diagrams can look different in the HID, the sequence of states for the whole outburst generally follows as: LHS $\rightarrow$ HIMS $\rightarrow$ SIMS $\rightarrow$ HSS $\rightarrow$ SIMS $\rightarrow$ HIMS $\rightarrow$ LHS (see \citealt{2011MNRAS.410..679M,2016ASSL..440...61B} for a review).

Quasi-periodic oscillations (QPOs) can be observed in the light curves of X-ray binary systems during outbursts. QPOs are studied mainly in the Fourier domain and are found in the power spectrum, which is the squared modulus of the Fourier transform of the light curve \citep{1989ASIC..262...27V}. In black hole X-ray binaries, QPOs with centroid frequency $\lesssim$ 30 Hz are defined as low frequency (LF) QPOs, and QPOs with centroid frequency $\gtrsim$ 60 Hz are defined as high frequency (HF) QPOs (see, e.g., \citealt{2010LNP...794...53B}). HF QPOs are scarce and only a few candidates are observed, e.g., XTE J1550-564 \citep{2002ApJ...580.1030R}, XTE J1650-500 \citep{2003ApJ...586.1262H}, GRO J1655-40 \citep{1999ApJ...522..397R,2002ApJ...580.1030R}, GRS 1915+105 \citep{1997ApJ...482..993M,2001ApJ...554L.169S,2006MNRAS.369..305B}. LF QPOs are frequently observed in BH X-ray binaries with a high signal-to-noise, which allows detailed studies on the properties and origins of QPOs in the past few decades. 

LF QPOs in BH X-ray binaries are generally classified into three types: A, B, and C (see \citealt{1999ApJ...526L..33W,2005ApJ...629..403C,2015MNRAS.447.2059M,2019NewAR..8501524I} for a review). Type-A QPOs are the least common LF QPOs in black hole X-ray binaries, e.g., GX 339-4 \citep{2011MNRAS.418.2292M}, XTE J1817-330 \citep{2012A&A...541A...6S}, XTE J1859+226 \citep{2013ApJ...775...28S}. Type-A QPOs are weak (few percent rms) and broad peaks in the power density spectra (PDS) with a centroid frequency of approximately 6-8 Hz in the power spectrum. Type-B QPOs have been detected in a large number of black hole X-ray binaries, e.g., GX 339-4 \citep{2011MNRAS.418.2292M}, XTE J1817-330 \citep{2012A&A...541A...6S}, GRO J1655-40 \citep{2012MNRAS.427..595M}, XTE J1859+226 \citep{2013ApJ...775...28S}, 4U 1543-47 \citep{2020MNRAS.495..182R}. Type-B QPOs with centroid frequencies typically at $\sim 5-6$ Hz, have a relatively high amplitude (up to $\sim$ 5\% rms) and are narrow (Q $\gtrsim$ 6) peaks. Type-C QPOs are the most common type of QPOs in BH systems. Type-C QPOs are characterized by a high amplitude (up to 20\% rms) and a narrow peak, with a wide frequency covering $\sim 0.1-30 $ Hz. In the PDS of type-C QPOs, a broadband, flat-topped noise can also be observed, which may be caused by fluctuations of the accretion rate. The properties of LF QPOs are related to their spectral states. Type-C QPOs are usually observed in the hard state and hard intermediate state. Type-B and Type-A QPOs are generally found in the soft intermediate state. Transitions between different types of QPOs can sometimes be observed during state transitions. (see, e.g., \citealt{2013ApJ...775...28S,2021MNRAS.505.3823Z}).

In recent years, many theoretical models for the production of Type-C QPOs have been proposed. These models generally can be classified into two types: geometrical, the shape or size of something around the black holes varies quasi-periodically, such as the relativistic precession model (RPM) (see, e.g., \citealt{1998ApJ...492L..59S,1999ApJ...524L..63S,2006ApJ...642..420S}), the precessing inner flow model \citep{2009MNRAS.397L.101I}, the corrugation modes model \citep{1980PASJ...32..377K,1999PhR...311..259W,2001PASJ...53....1K}, the accretion ejection instability model \citep{1999A&A...349.1003T}, the propagating oscillatory shock model \citep{1996ApJ...457..805M,2008A&A...489L..41C}; intrinsic, assuming the geometry of the system to be stable, with some resonant oscillations in properties of the disk, like density, pressure or accretion rate. 

In recent years, some studies have provided evidence that Type-C QPOs may have a geometric origin. \citet{2015MNRAS.447.2059M} and \citet{2015MNRAS.448.3348H} statistically showed that Type-C QPOs are stronger in higher inclination binary systems. \cite{2017MNRAS.464.2643V} showed that small hard lags (i.e., hard photons lag soft photons) are seen when the Type-C QPO has a low frequency, and hard/soft lags are observed in low/high inclination sources when the Type-C QPO has a high frequency. \citet{2016MNRAS.461.1967I} showed that the centroid energy of the iron K$\alpha$ line varies with Type-C QPO phase. Therefore, reflection features (especially the iron line) contain information around the BH and its accretion disk, which may be able to provide valuable diagnostics for the physical origin of Type-C QPOs.

\begin{table*}[!htbp]
\renewcommand{\arraystretch}{1.5}
    \centering
    \caption{\textit{Insight}-HXMT observations of 4U 1630-47 in the 2021 outburst with properties of detected QPOs and QRMs (measured from the PDS in ME bands of 10-35 keV). The letter P indicates that the error of the parameter was pegged at the upper or lower boundary. All errors are calculated at 90 percent confidence level.}
    \begin{tabular}{cccccccccc}
       \hline
            &                &     &     &  \multicolumn{3}{c}{QPO/QRM}    & \multicolumn{2}{c}{Broadband noise} & \\ \cline{5-7} \cline{8-9}
       Num. & Observation ID & MJD & PDS & Centroid & Fractional & Q factor & Centroid& Fractional & $\chi^2$/d.o.f \\  
        &  &  & & Frequency (Hz) &  rms (\%) & & Frequency (Hz) &  rms (\%)& \\ 
       \hline
       1 & P040426300101  &   59475.647  &     QPO       &  $1.66^{+0.05}_{-0.05}$     &  $17.8^{+2.8}_{-3.8}$  &  $4.9_{-1.6}^{+0.8}$  &  $0.0^{+2.1}_{-P}$ &  $12.2^{+5.4}_{-8.4}$  & 140/117  \\
       2 & P040426300201  &   59476.488  &     QPO       &  $2.43^{+0.03}_{-0.03}$     &  $15.3^{+1.3}_{-1.1}$  &  $8.2_{-0.2}^{+0.2}$  &  $0.0^{+0.2}_{-P}$ &  $17.5^{+1.7}_{-1.5}$  & 146/117   \\
       3 & P040426300301  &   59477.647  &     QPO       &  $2.91^{+0.08}_{-0.07}$     &  $13.0^{+2.8}_{-1.8}$  & $11.1_{-1.0}^{+0.4}$  &  $0.0^{+0.3}_{-P}$ &  $21.4^{+2.7}_{-2.5}$  & 133/117\\
       4 & P040426300401  &   59478.906  &     QPO       &  $3.54^{+0.08}_{-0.07}$     &  $15.1^{+2.1}_{-1.1}$  &  $6.3_{-0.3}^{+0.4}$  &  $0.0^{+0.1}_{-P}$ &  $24.4^{+1.2}_{-1.3}$  & 131/117\\
       5 & P040426300501  &   59479.647  &    None      &  None                       &  None               &  None &  $0.0^{+0.1}_{-P}$ &  $25.4^{+1.0}_{-1.1}$   &  165/117 \\
       6 & P040426300601  &   59480.641  &    None      &  None                    &  None               &  None & $0.0^{+0.1}_{-P}$  &    $23.9^{+1.1}_{-1.2}$      &  148/117  \\       
       7 & P040426300701  &   59481.667  &     QPO       &  $2.39^{+0.07}_{-0.08}$     &  $14.5^{+3.3}_{-4.7}$  &  $7.0_{-0.8}^{+0.7}$  &  $0.0^{+1.8}_{-P}$ &  $19.4^{+0.5}_{-0.7}$  & 141/117\\
       8 & P040426300801  &   59482.626  &     QPO       &  $3.27^{+0.11}_{-0.14}$     &  $16.4^{+2.5}_{-2.3}$  &  $3.9_{-0.4}^{+0.6}$  &  $0.0^{+0.2}_{-P}$ &  $23.0^{+0.2}_{-0.2}$  & 128/117\\
       9 & P040426300901  &   59483.785  &     QRM       &  $0.066^{+0.005}_{-0.003}$  &  $12.3^{+1.2}_{-1.0}$  &  $2.2_{-0.3}^{+0.5}$  &  $0.0^{+0.3}_{-P}$ &  $12.9^{+0.9}_{-1.1}$  & 288/218\\
       10 & P040426300902  &   59483.888  &    QRM       &  $0.066^{+0.002}_{-0.002}$  &  $14.0^{+0.8}_{-0.8}$  &  $3.5_{-0.2}^{+0.2}$  &  $0.0^{+0.1}_{-P}$ &  $11.3^{+0.8}_{-0.8}$  & 264/218\\
       11 & P040426300903  &   59484.022  &    QRM       &  $0.063^{+0.009}_{-0.004}$  &  $10.5^{+1.0}_{-0.8}$  &  $1.5_{-0.4}^{+0.8}$  &  $0.0^{+0.7}_{-P}$ &  $13.5^{+0.9}_{-1.7}$  & 240/218\\
       12 & P040426301001  &   59484.932  &    QRM       &  $0.054^{+0.003}_{-0.003}$  &  $14.0^{+0.9}_{-0.8}$  &  $1.7_{-0.2}^{+0.2}$  &  $0.0^{+0.3}_{-P}$ &  $12.8^{+0.9}_{-0.8}$  & 239/218\\
       13 & P040426301002  &   59485.050  &    QRM       &  $0.052^{+0.003}_{-0.001}$  &  $15.8^{+1.4}_{-1.6}$  &  $1.9_{-0.1}^{+0.3}$  &  $0.0^{+0.2}_{-P}$ &  $9.5^{+1.0}_{-1.2}$  & 249/218\\
       14 & P040426301003  &   59485.240  &    QRM       &  $0.056^{+0.002}_{-0.004}$  &  $10.9^{+1.1}_{-1.2}$   &  $3.5_{-0.3}^{+0.7}$  &  $0.0^{+0.2}_{-P}$ &  $8.4^{+0.5}_{-1.1}$  & 224/218\\
       15 & P040426301004  &   59485.373  &    None      &  None                       &  None                     &  None &  $0.0^{+0.3}_{-P}$   &  $4.1^{+1.4}_{-1.4}$   &  125/117\\
       16 & P040426301005  &   59485.506  &    None      &  None                       &  None                     &  None &  $0.0^{+0.1}_{-P}$   &  $2.1^{+2.2}_{-0.7}$   &  138/117\\
       \hline
    \end{tabular}
    \label{table1}
\end{table*}

In addition to QPOs, oscillations with a period of $\sim$10-200 s have been observed in the light curve of some black hole candidates, e.g., 4U 1630-47 \citep{2001MNRAS.322..309T}, GRS 1915+015 \citep{1997ApJ...482..993M}, GRO J1655-40 \citep{1999ApJ...522..397R}, IGR J17091-3624 \citep{2011ApJ...742L..17A}, and H1743-322 \citep{2012ApJ...754L..23A}. These variability patterns generally have a frequency of several tens of mHz and appear as a broad peak in the PDS. \citet{2001MNRAS.322..309T} named these quasi-regular flares observed in 4U 1630-47 as quasi-regular modulations (QRMs). The most focused variability class of long-time quasi-regular flares or dips is the $“\rho”$ class (or “heartbeat” state). It's characterized by regular X-ray flares typically lasting $\sim 40-200$ s (see, e.g., \citealt{2018ApJ...865...19W}). A possible origin of it is radiation-pressure-driven evaporation or ejections that happen at the inner region of the disk \citep{2012ApJ...750...71N}. The radiation pressure will be dominant at the inner region of the accretion disk, and when the mass accretion rate reaches a certain level, the limit-cycle behavior of the disk can lead to the modulation of flux due to the thermal viscous instability \citep{1974ApJ...187L...1L}. Tight correlations among the recurrence time, the inner radius of the disk, and the luminosity of the non-thermal emission have been found in the analysis of the “heartbeat” state \citep{1997ApJ...482..993M}. And \citet{2018ApJ...865...19W} suggested that the “heartbeat” state in GRS 1915+105 may be caused by the variation of the corona size.


The spectra of BH X-ray binaries are generally composed of a black body component and a power-law component. The thermal black body component is generally considered to originate from a geometrically thin, optically thick accretion disk \citep{1973A&A....24..337S,1973blho.conf..343N}. The turbulent stress will transport the angular momentum outward, and the disk material falls into the black hole releasing its gravitational potential energy. The inner disk will be heated and emit thermal emission. The accretion disk may evaporate and float into a large-scale height and play the role of the corona \citep{1975ApJ...199L.153E,1977ApJ...214..840I}. Compton up-scattering of soft X-ray photons from the disk by the corona is commonly considered the origin of the power-law component \citep{1975ApJ...195L.101T,1979Natur.279..506S}. The spectrum may also display a reflection component coming from the accretion disk reflecting the coronal emission. The reflected emission will be reprocessed by the disk’s upper atmosphere and re-emit with characteristic features such as an iron K$\alpha$ fluorescence line at $\sim$ 6.4 - 6.97 keV, depending on the ionization of the iron ions, and a broad bump peaking at $\sim$ 20 - 30 keV which is named as the Compton hump \citep{1980ApJ...236..928L,2010ApJ...718..695G}. The reflection features near the BH will be distorted by the relativistic motion of disk material and the gravity of the BH. As a result, the iron line profile appears asymmetrically broadened in the spectrum \citep{1989MNRAS.238..729F,2010MNRAS.409.1534D}. 

Black hole X-ray binary 4U 1630-47 is a transient source showing recurrent outbursts. It was first observed by Vela-5B during the 1969 outburst \citep{1986Ap&SS.126...89P}. Since then, 4U 1630-47 goes into outbursts frequently and with a typical period of 600-700 days \citep{1976ApJ...210L...9J,1986Ap&SS.126...89P,1995ApJ...452L.129P}, and it is a part of the group of X-ray transients known as X-ray novae \citep{1994AstL...20..777S,1996ARA&A..34..607T}. X-ray novae are assumed to be recurrent, with a typical outburst recurrence time of about 10-50 years. Therefore, only a single event is recorded in some sources \citep[e.g.,][]{1997ApJ...491..312C,2014AAS...22340606G}. The source 4U 1630-47 lies in the direction toward the Galactic center and is heavily absorbed in the soft X-ray band, which implies a large distance to the source. Due to the large extinction amount of optical ($\textgreater$20 mag) and reddening effect, no optical counterpart has been identified and no dynamical mass measurement has been made at present \citep{2014ApJ...789...57S}. \citet{2014ApJ...789...57S} studied the correlation between the mass accretion rate and the photon index of spectra (observations that LF QPOs are observed), estimated 4U 1630-47 has a mass $\sim10 M_{\sun}$ and an inclination angle $i \leq70^{\circ}$. \citet{2018ApJ...859...88K} studied the dust-scattering halo created by 4U 1630-47 and estimated the distance to the source to be about 4.7-11.5 kpc. \citet{2014ApJ...784L...2K} estimated an extremely high spin $a = 0.985^{+0.005}_{-0.014}$ and an inclination angle $i = 64_{-3}^{+2}$ deg by fitting the NuSTAR spectra. \citet{2022MNRAS.512.2082L} obtained a moderately high spin $a = 0.817\pm 0.014$ from fitting the reflection spectra observed by \textit{Insight-}HXMT during the 2020 outburst.

\begin{figure}[!htbp]
   \centering
   \includegraphics[width=\columnwidth]{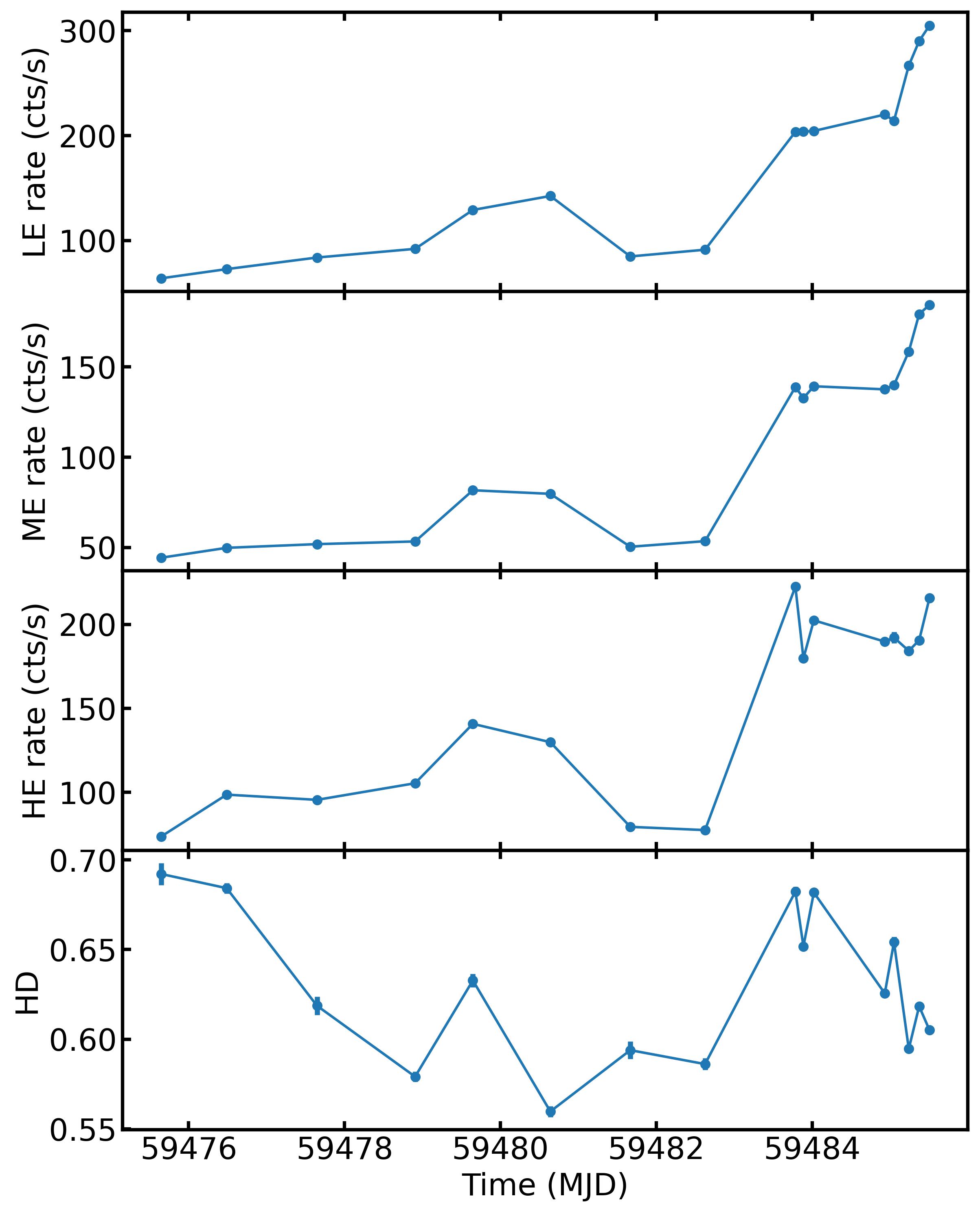}
   \caption{Insight-HXMT LE (2-10 keV), ME (10-35 keV), and HE (27-100keV) light curves of the source, together with the variation of the hardness ratio defined as the ratio of the count rates between the ME 10-35 keV and LE 2-10 keV bands. The background rates are generally lower than the count rates of the source in corresponding energy bands.}
   \label{figure1}
\end{figure}

In this work, we studied the quasi-periodic oscillations and spectral evolutions of the 4U 1630-47 based on \textit{Insight}-HXMT observations of its 2021 outburst. In Section 2, we introduce the observations and data reduction process. The timing analysis on QPOs and QRMs and spectral results are shown in Section 3. We briefly discuss our results and implications in Section 4. A concise conclusion of the work is given in section 5.

\section{Observations and data reduction}
The Hard X-ray Modulation Telescope \textit{Insight}-HXMT is China’s first X-ray astronomy satellite, launched on 2017 June 15 \citep{2020SCPMA..6349502Z}. It is a large X-ray astronomical satellite with a broad energy band of 1-250 keV. To fulfill the requirements of the broadband spectra and fast variability observations, three payloads are configured onboard \textit{Insight}-HXMT: High Energy X-ray telescope (HE) for 20-250 keV band \citep{liu2020high}, Medium Energy X-ray telescope (ME) for 5-30 keV band \citep{cao2020medium}, and Low Energy X-ray telescope (LE) 1-15 keV band \citep{chen2020low}. Light curves and spectra were extracted using \textit{Insight}-HXMT Data Analysis Software (HXMTDAS) v2.05 following the standard procedure (also see processing details described in \citealt{WANG20211,chen2021relation}). In the data screening procedure, we use tasks $he/me/lepical$ to remove spike events generated by electronic systems and $he/me/legtigen$ to select good time interval (GTI) under the following conditions: the pointing offset angle $< 0.04^\circ$; the pointing direction above earth $> 10^\circ$; the geomagnetic cut-off rigidity > 8 GeV and no South Atlantic Anomaly (SAA) passage occurred within the previous 300 seconds. Background estimation was performed using \textit{he/me/lebkgmap}, and further refined using \textit{lcmath} to eliminate estimated background noise. The dead-time effect is corrected using the nonparalyzable model proposed in \citet{1995ApJ...449..930Z}.

\textit{Insight-}HXMT started high-cadence monitoring of 4U 1630-47 from 2021 September 18 to 28 and stopped after 2021 September 28 owing to the small solar aspect angle ($\text{\textless}70^{\circ}$, \citealt{2022ApJ...937...33Y}). The observations we studied in this work are listed in Table \ref{table1}, they are observed by \textit{Insight-}HXMT. These observations are in the hard state or hard intermediate state and display reflection features in the spectra. Therefore, we fit the spectra with the reflection model and study the reflection features. In this work, we analyzed the spectra using 2-8 keV for LE, 8-28 keV for ME, and 28-100 keV for HE.

\begin{figure*}[!htbp]
   \centering
   \includegraphics[width=0.7\columnwidth]{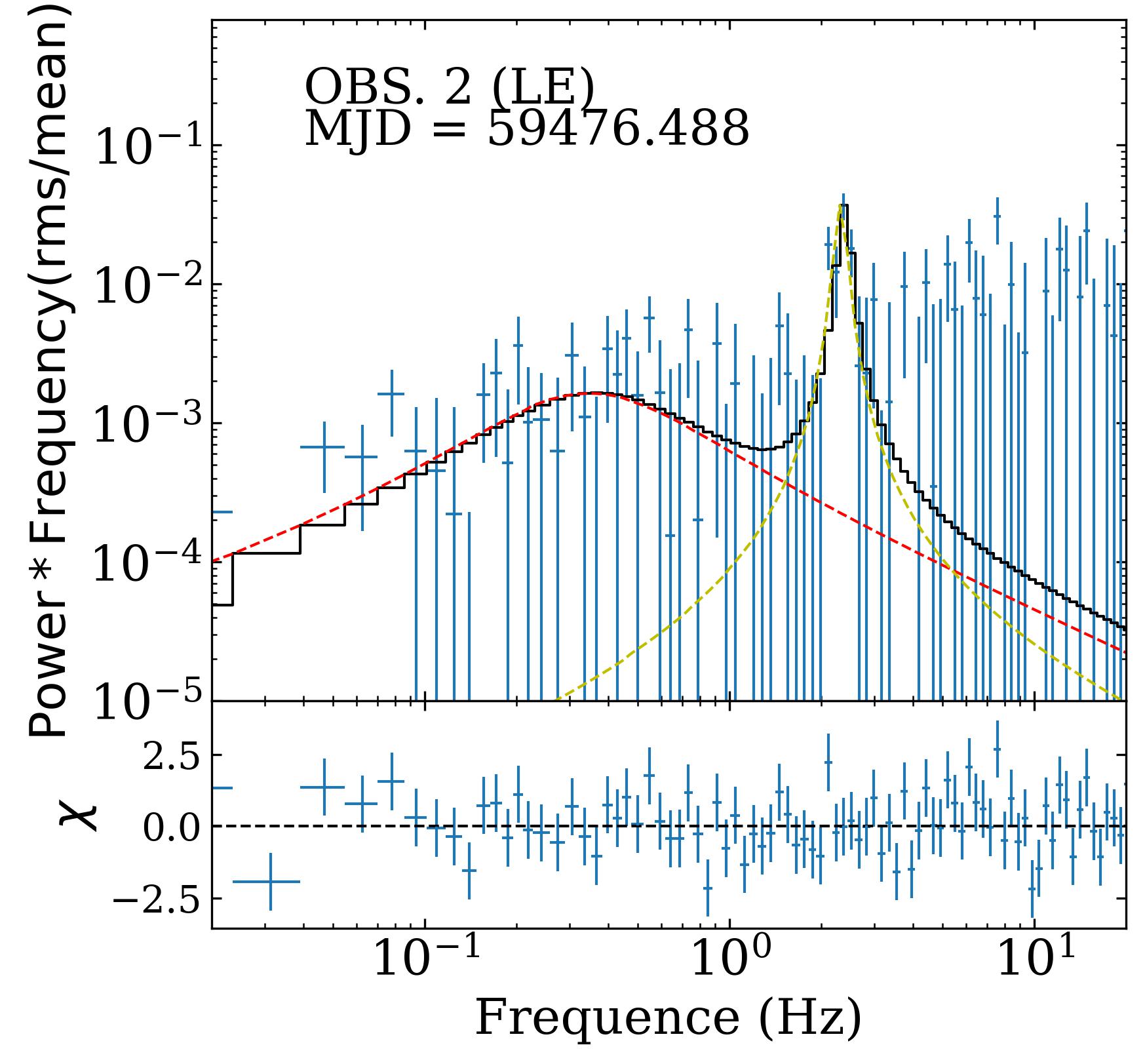}
   \includegraphics[width=0.7\columnwidth]{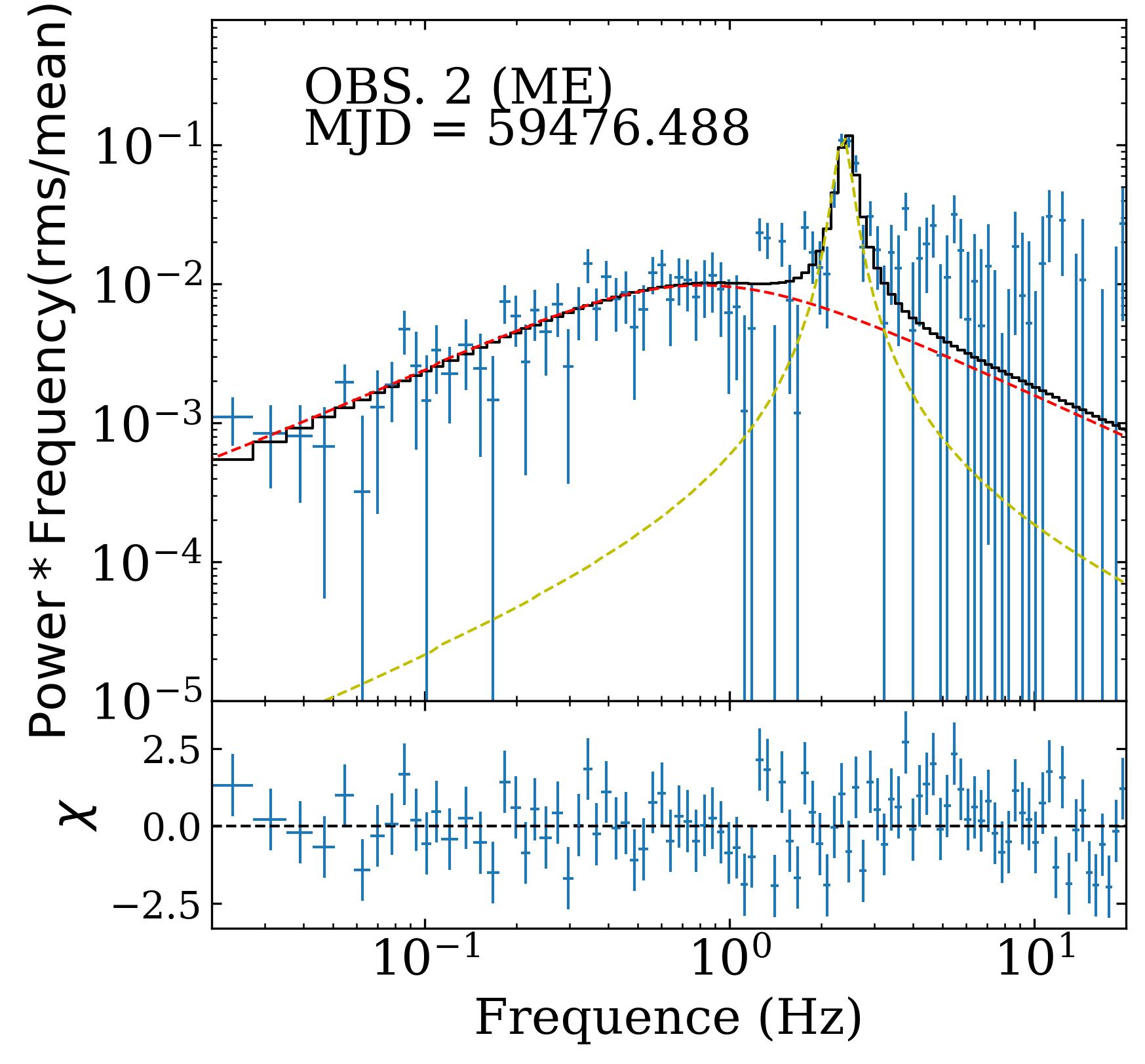}
   \includegraphics[width=0.7\columnwidth]{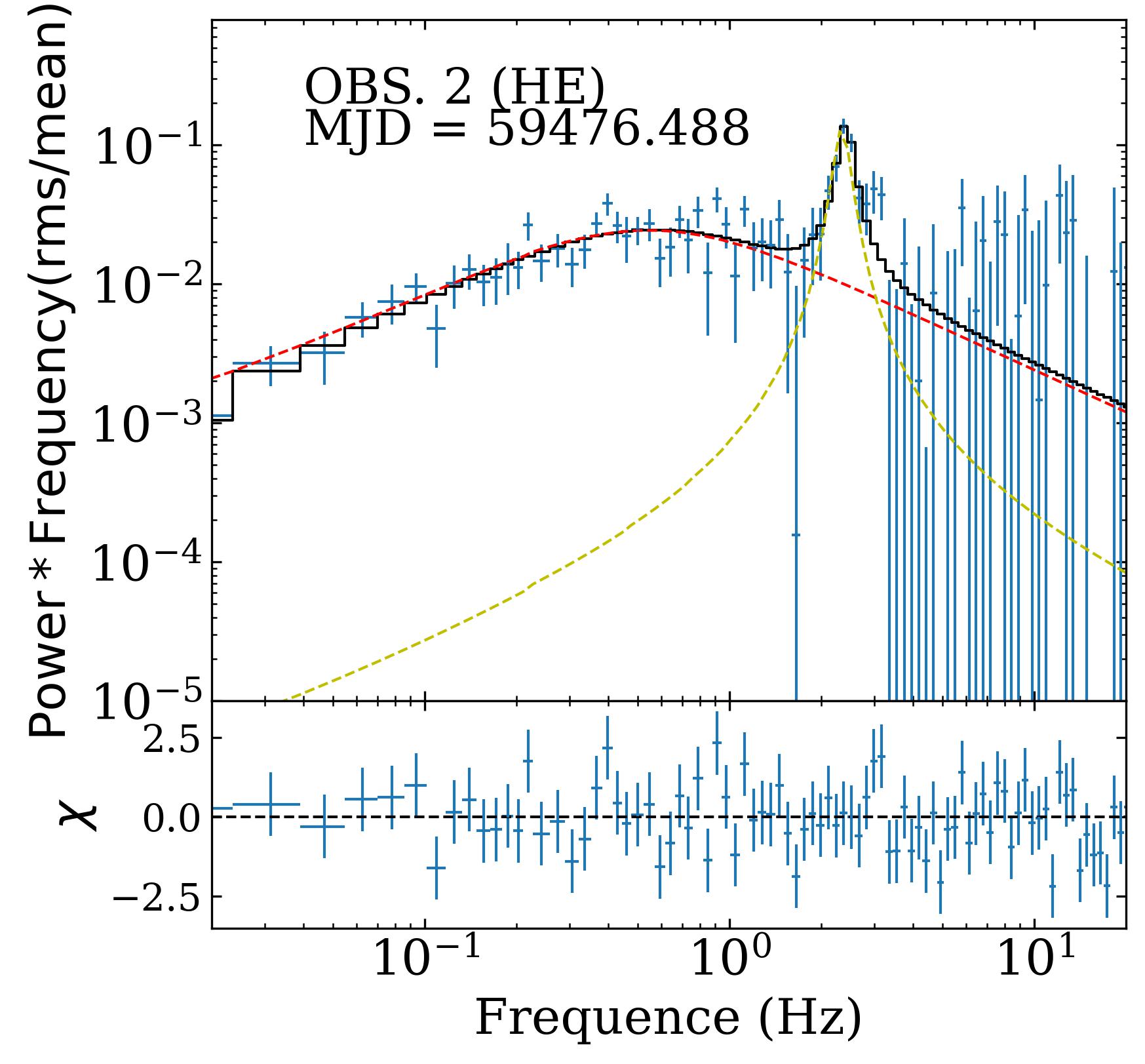}
   \caption{The representative fitting results of the QPO signals observed in 4U 1630-47 with Insight-HXMT. \textbf{Left:} PDS of OBS. 2 calculated from the LE band (2 - 10 keV), the QPO signal is weaker with the rms $\sim7\%$. \textbf{Middle:} PDS of OBS. 2 calculated from the ME band (10 - 35 keV) shows the QPO feature around 2.5 Hz. \textbf{Right:} PDS of OBS. 2 calculated from the HE band (27 - 100 keV), the properties of QPO are similar to the ME band.}
   \label{figure2}
\end{figure*}

\begin{figure}[!htbp]
   \centering
   \includegraphics[width=0.9\columnwidth]{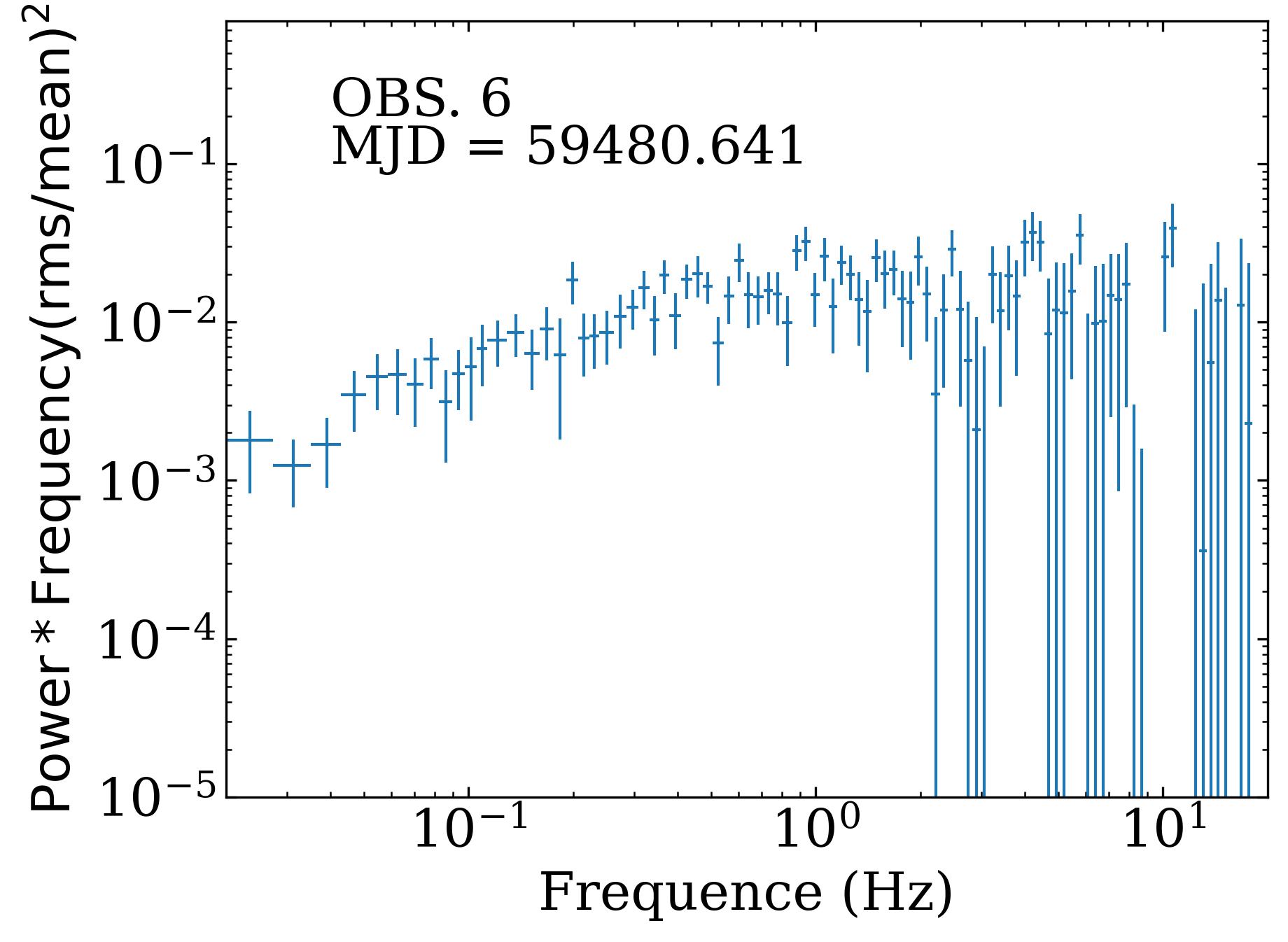}
   \caption{The representative PDS (ME band, 10 - 35 keV) of no evident QPO signals.}
   \label{figurenone}
\end{figure}

\begin{figure}[!htbp]
   \centering
   \includegraphics[width=\columnwidth]{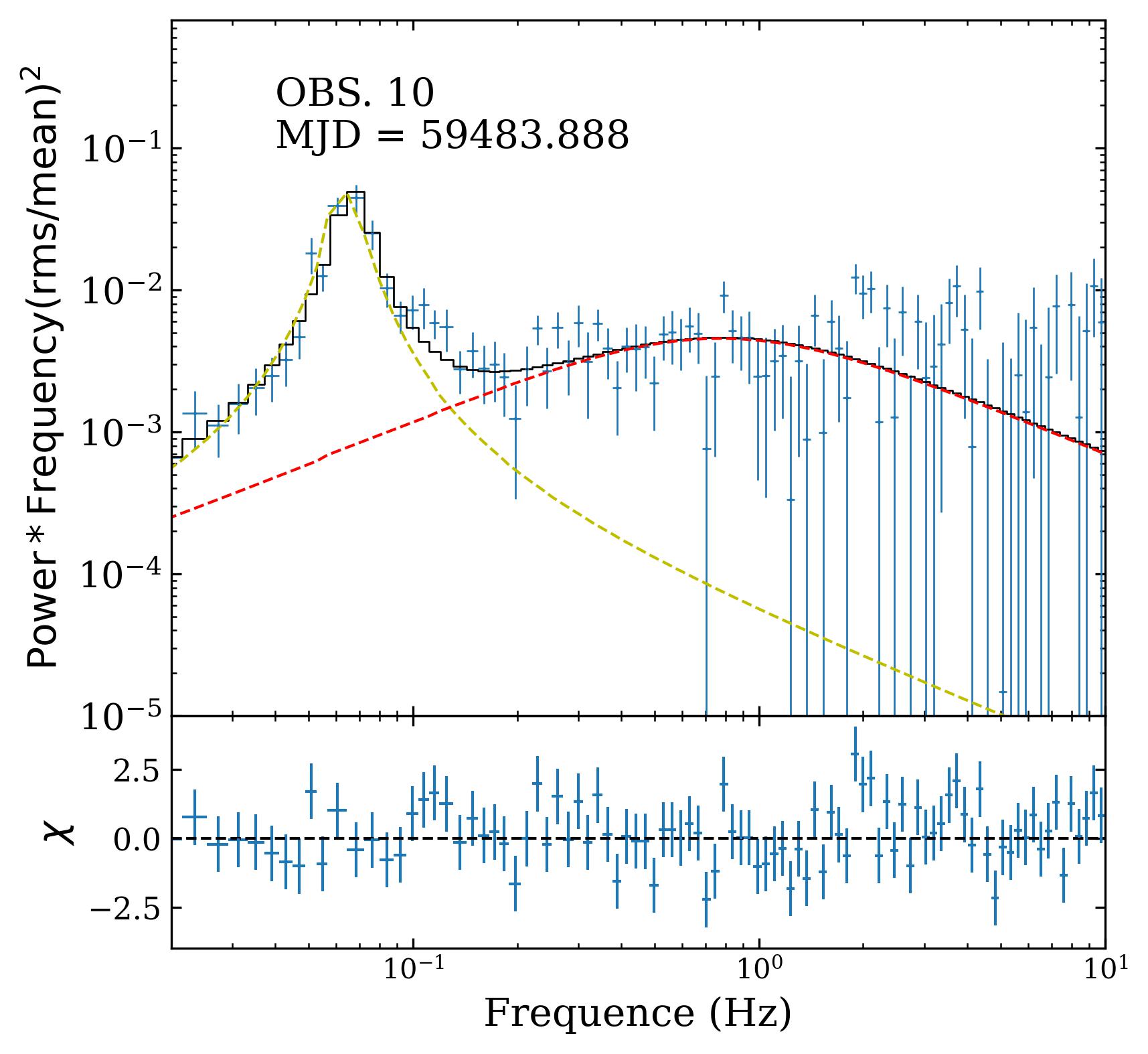}
   \caption{The representative fitting result of the QRM signals observed in 4U 1630-47 with Insight-HXMT. The figure is the PDS of OBS. 10 calculated from the ME band, and shows the QRM feature around 0.06 Hz.}
   \label{figure6}
\end{figure}

\begin{figure*}[!htbp]
   \centering
   \includegraphics[width=0.63\columnwidth]{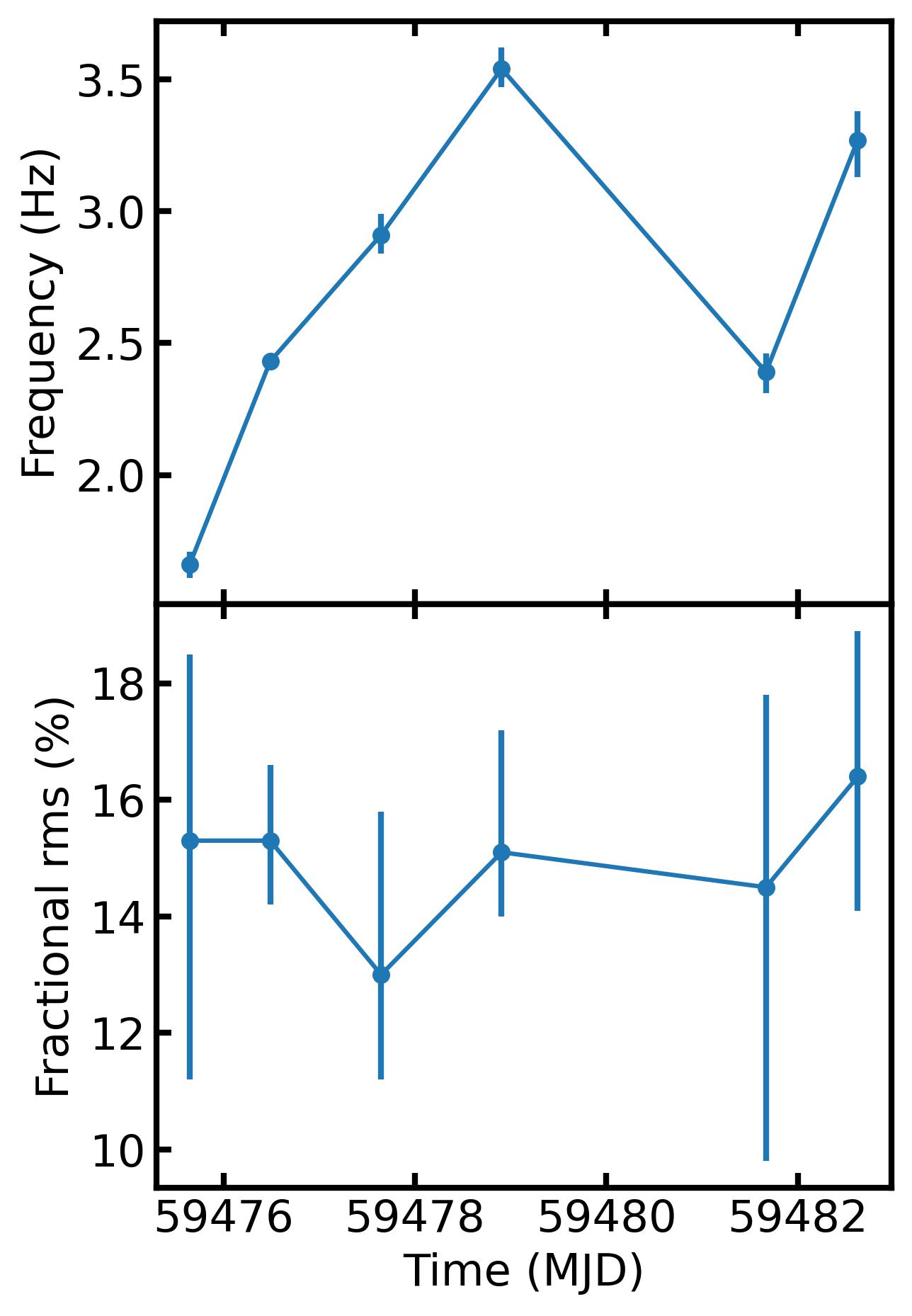}
   \includegraphics[width=0.67\columnwidth]{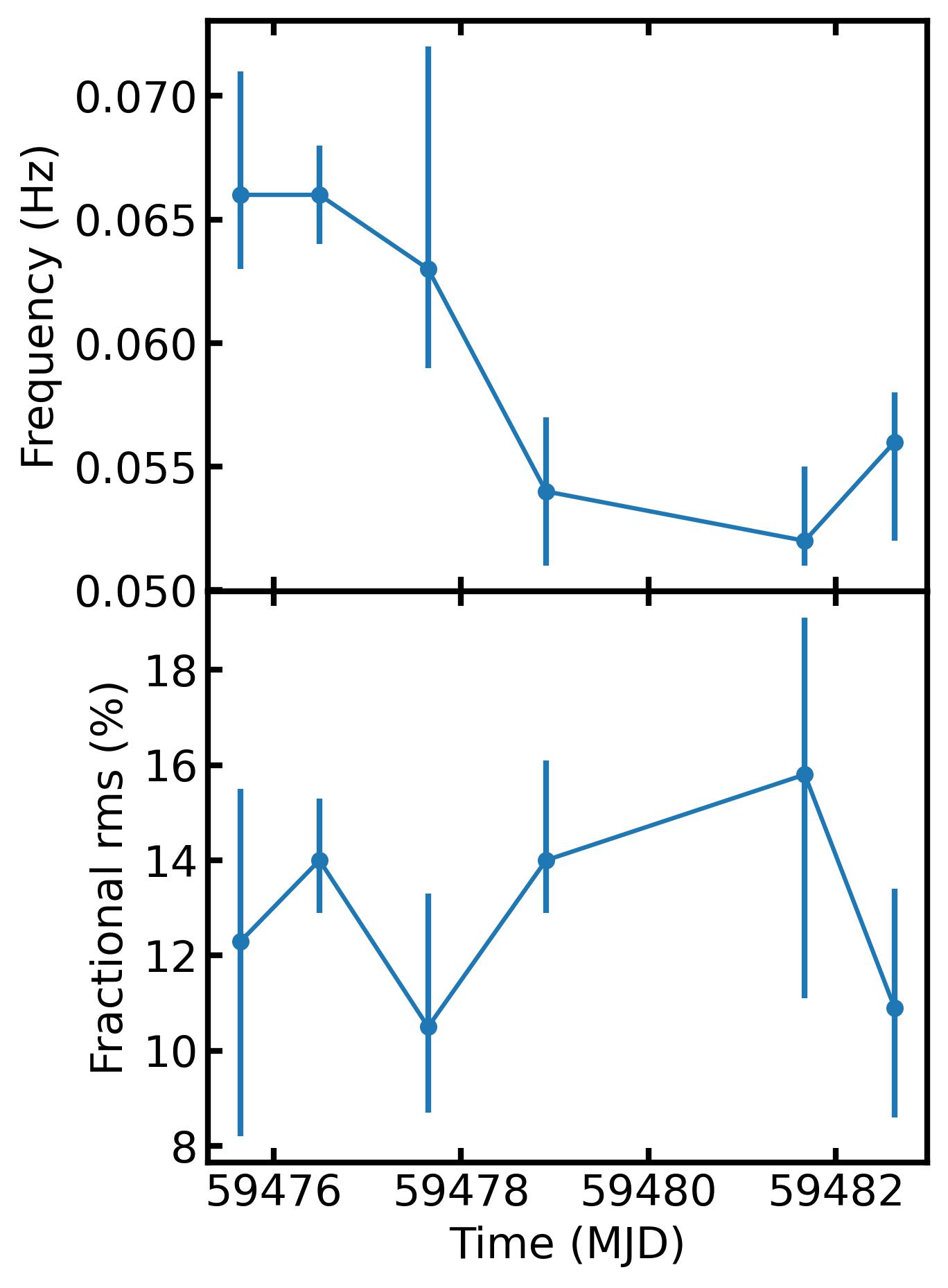}
   \caption{The evolution of the frequency and fractional rms for both QPOs and QRMs. \textbf{Left:} the evolution of QPOs. \textbf{Right:} the evolution of QRMs.}
   \label{figure1p}
\end{figure*}

We employ powspec from HEASOFT to calculate the PDS for each observation to obtain the properties of QPOs and QRMs. We use a time interval of 128 s and a corresponding time resolution of 1/128 s for the QPO signals and a time interval of 256 s and a corresponding time resolution of 1/64 s for the QRM signals. The PDS are rebinned in frequency space using a geometric factor of 1.05. The PDSs were normalized to units of rms$^2$Hz$^{-1}$ and subtracted Poisson noise following the methods described in \citet{1990A&A...230..103B} and \citet{1991ApJ...383..784M}. We utilized multiple Lorentz functions to fit the profiles of the broadband noises and QPOs/QRMs in the PDS. This approach allows us to extract the fundamental parameters of the QPOs/QRMs. The fractional rms of the QPOs/QRMs is calculated:
\begin{equation}
    rms_{\rm{QPO}}=\sqrt{R}\times\frac{S+B}{S},
\end{equation}
where S is the source count rate, R is the normalization of the Lorentzian component, and B is the background count rate (see, e.g., \citealt{2015ApJ...799....2B,2024ApJ...968..106Z}). And obtain the quality factor $Q=\nu/\Delta\nu$ (where $\nu$ represents the frequency of the QPO and $\Delta\nu$ represents the full width at half maximum, FWHM). 

\begin{figure*}
   \centering
   \includegraphics[width=\columnwidth]{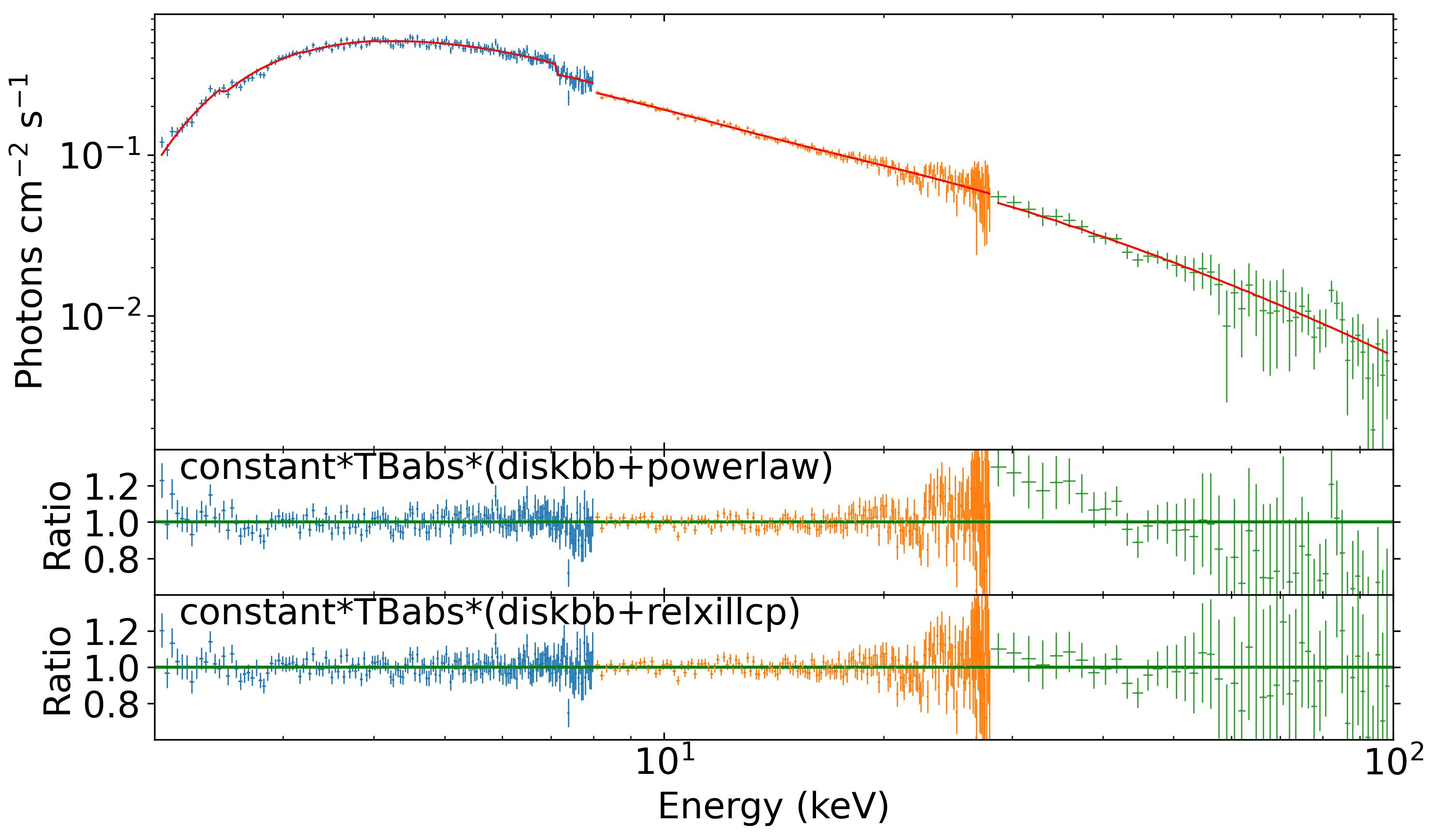}
   \includegraphics[width=\columnwidth]{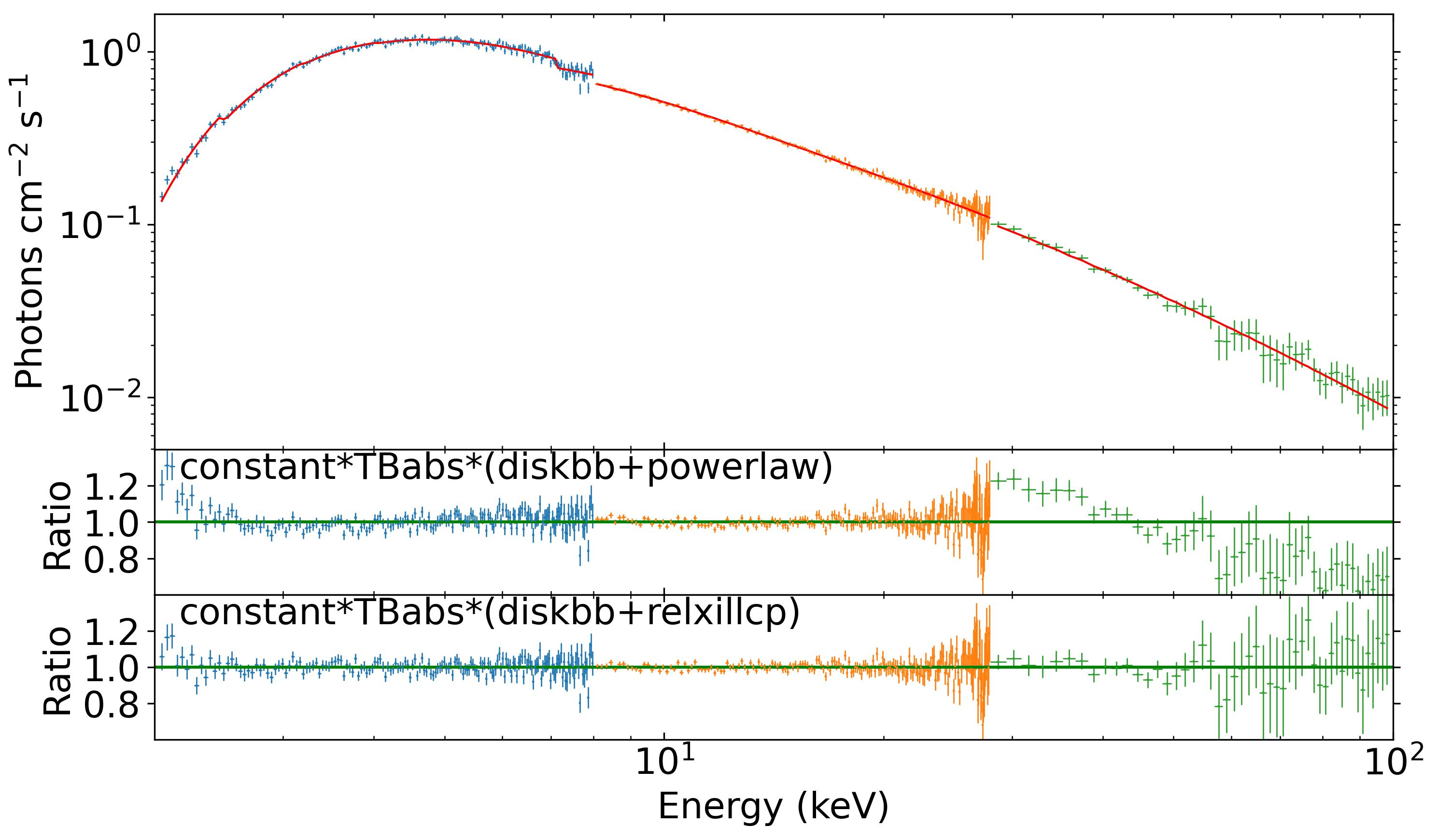}
   \caption{The best-fitting results and the corresponding data-to-model ratios of the three spectra using LE, ME, and HE detectors. The spectra are fitted with model \texttt{constant*tbabs(diskbb+relxillcp)}. \textbf{Top:} Observation Num. 4, when QPO is detected. \textbf{Bottom:} Observation Num. 10, when QRM is detected.}
   \label{figure3}
\end{figure*}

\section{analysis and results}

Figure \ref{figure1} presents the background-subtracted light curves from Insight-HXMT in three energy bands (LE, ME, and HE) and the hardness ratios (count ratio of ME to LE). We create an averaged power spectrum for each observation to study the fast X-ray variability. Examples of the PDS are shown in Figures \ref{figure2}, \ref{figurenone}, and \ref{figure6}, accompanied by respective model fitting results. The evolution of both frequency and fractional rms of QPOs and QRMs is shown in Figure \ref{figure1p}. The QPOs in HE and LE bands of OBS. 2 have similar frequencies to that in the ME band, but the signal is weaker (fractional rms $\sim7\%$) in the LE band. Apart from Obs. 2, the PDSs of other observations also show very weak or no QPO signal in LE and HE bands. Therefore, we use ME band light curves to calculate the PDS and QPO properties in the following. The fitting results of all QPOs and QRMs are listed in Table \ref{table1}. From MJD 59475.6-59482.6 (OBS. 1-4, 7-8), the PDS features a relatively narrow QPO peak centered at 1.6-3.6 Hz, accompanied by a broadband noise. The frequency of the QPOs is anti-correlated with the spectral hardness ratio. The Q factors are in the range of $\sim 4-11$, with fractional rms amplitudes of $\sim12{\%}-17\%$. Based on these characteristics, these QPOs can be identified as Type-C QPOs (see, e.g., \citealt{1999ApJ...526L..33W,2005ApJ...629..403C}). From MJD 59483.8-59485.2 (OBS. 9-14), the PDS features a relatively broad peak near 60 mHz (named QRM) along with a broadband noise component. The QPOs are observed when the flux is low and the QRMs are observed in a certain range of flux (e.g., ME 10-35 keV count rates $\sim$ 120 - 150 counts s$^{-1}$). 

During the Insight-HXMT observations, the source was located in the hard and intermediate states (see Figure 2 in \citealt{2022ApJ...937...33Y}), which would be suitable to study non-thermal spectral properties, e.g., reflection features. We used spectral analysis software Xspec v12.14.0 to study the spectra \citep{1996ASPC..101...17A}. Firstly, we fit the spectra with a phenomenological model \texttt{constant*tbabs(diskbb+powerlaw)}. Figure. \ref{figure3} shows two examples of spectral fitting results, OBS. 4 (QPO) and 10 (QRM). The spectra show a Compton hump $\sim$ 30 keV which is attributed to the reflection component. The fitting of OBS. 4 returns a $\chi^2_{\upsilon}=1.01$, and OBS. 10 returns $\chi^2_{\upsilon}=1.61$. Then, we fit the spectra with a combined model \texttt{constant*tbabs(diskbb+relxillcp)}. Model \texttt{relxillcp} is a physical model for relativistic reflection with the \texttt{nthcomp} model as the primary source spectrum \citep{2014ApJ...782...76G,2014MNRAS.444L.100D}. This improves $\chi^2_{\upsilon}$ to 0.92 and 1.00, respectively, thus replacing \texttt{powerlaw} by \texttt{relxillcp} improves the fitting results. Besides, we tried to fit Obs. 10 with other reflection models \texttt{relxilllp} and \texttt{relxilllpcp}. The \texttt{relxilllp} model assumes that the corona is a point source located at a height above the central compact object. Most of the parameters in the \texttt{relxilllp} model are the same as in \texttt{relxill}. But instead of the emissivity index, \texttt{relxilllp} has two new parameters, $h$ and $\beta$, which are the height of the corona and the corona velocity. In the \texttt{relxilllpCp} model, the accretion disk is illuminated by a thermal Comptonization spectrum instead of a power-law spectrum in the \texttt{relxilllp} model. The fitting returns $\chi^2_{\upsilon}=1.47$ and $\chi^2_{\upsilon}=1.45$. 

Usually, if the normalization of LE is fixed to 1, the normalization of ME and HE should be close to 1. There are minor differences between the calibration of the two detectors due to the effects of systematic errors \citep{2020JHEAp..27...64L}. The relative differences may change slightly during the fitting process. Model \texttt{constant} is used for coordinating calibration differences between detectors. In the fitting, we fixed the normalization of LE to 1, the normalizations of ME and HE are set free in the range of 0.85 to 1.15. Model \texttt{tbabs} fits the galactic absorption \citep{2000ApJ...542..914W}. Due to the uncertain distance of 4U 1630-47, the hydrogen column density remains poorly constrained \citep[$N_{\text{H}}\sim 5\text{-}12 \times 10^{22}$ $\text{cm}^{-2}$,][]{2022ApJ...937...33Y}. Therefore, we set the galactic hydrogen column density $N_{\text{H}}$ free. The reflection model \texttt{relxillcp} is used to study the reflection features. We fix the density $\textrm{logN}=15$, and the electron temperature in the corona $\text{kT}_e=300$ keV since the fitting is insensitive to their values. The inclination angle of the accretion disk is fixed to 64$^{\circ}$, and the spin of BH is fixed to $a = 0.985$. The emissivity for the coronal flavor models $q_{in}$ and $q_{out}$ is linked. The free parameters in the model are the inner radius of the accretion disk ($R_{in}$), the emissivity index for the coronal flavor models ($q_{in}$ and $q_{out}$), power-law index of the primary source spectrum ($\Gamma$), iron abundance in solar units (${A_{\textrm{Fe}}}$), ionization of the accretion disk ($\log\xi$), reflection fraction parameter (${R_{f}}$).

The reflection fraction is defined as the ratio of the coronal intensity illuminating the disk to the coronal intensity that reaches the observer, independent of system parameters such as inclination \citep{2016A&A...590A..76D}. If there is no opaque material blocking the central emission region, the observer will see the sum of the radiation from the primary source and the reflected component. When the source height is reduced, the reflection fraction $R_{f}$ can increase markedly owing to the light-bending effects near the black hole \citep{2016A&A...590A..76D}. In addition, the black hole’s spin, $a$, modifies the area of the reflector (the inner edge of the accretion disk) and the amount of reflection. Thus, the reflection fraction can provide valuable insights into the geometry of the reflector \citep{2014MNRAS.444L.100D}.

The ionization parameter is defined by:
\begin{equation}
    \xi=4\pi F_\textrm{X}/n_\textrm{H},
\end{equation}
where $F_\textrm{X}$ is the flux of irradiation on the accretion disk, $n_\textrm{H}$ is disk hydrogen number density. The value of the ionization parameter ranges from 0 (neutral) to 4.7 (heavily ionized) in the model \texttt{relxillcp} and is sensitive to both the disk structure and the coronal illumination \citep{2011ApJ...734..112B,2014ApJ...782...76G}. The radial dependence of the irradiation will consequently lead to a radial dependence in terms of ionization as well.


The reflection emissivity is given by:
\begin{equation}
    \epsilon(r)=r^{-q},
\end{equation}
where $q$ is the emissivity index that can either be constant across all radii or vary with radius. For a point-like X-ray source at height $h$ on the disk axis, the disk irradiation is proportional to $(r^2+h^2)^{-3/2}\propto r^{3}$ in the absence of any relativistic effect \citep{1997ApJ...488..109R}.
Therefore, the emissivity profile with index $q=3$ is considered the standard. However, in the lamp-post geometry, the emissivity profile can vary significantly in the innermost regions depending on the position of the X-ray source \citep{2011MNRAS.414.1269W,2013MNRAS.430.1694D}. Assuming a constant density and ionization across the accretion disk, the emissivity index can reach high values when the source is near the black hole event horizon. Under such conditions, the emissivity profile can be very steep in the inner region of the disk; it flattens and finally reaches $q=3$ at a further radius \citep{2019MNRAS.485..239K}.

\begin{figure*}[!htbp]
   \centering
   \includegraphics[width=\columnwidth]{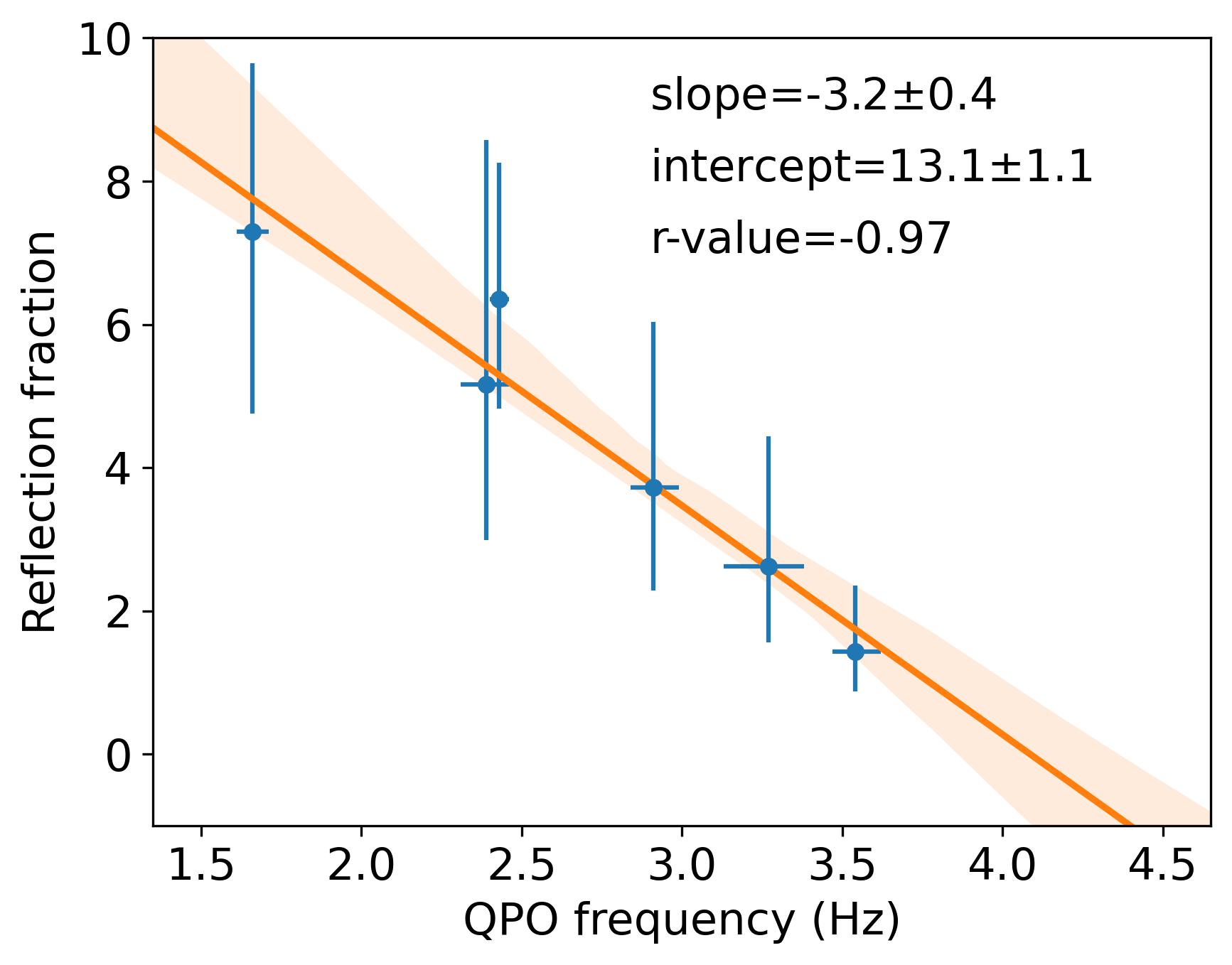}
   \includegraphics[width=\columnwidth]{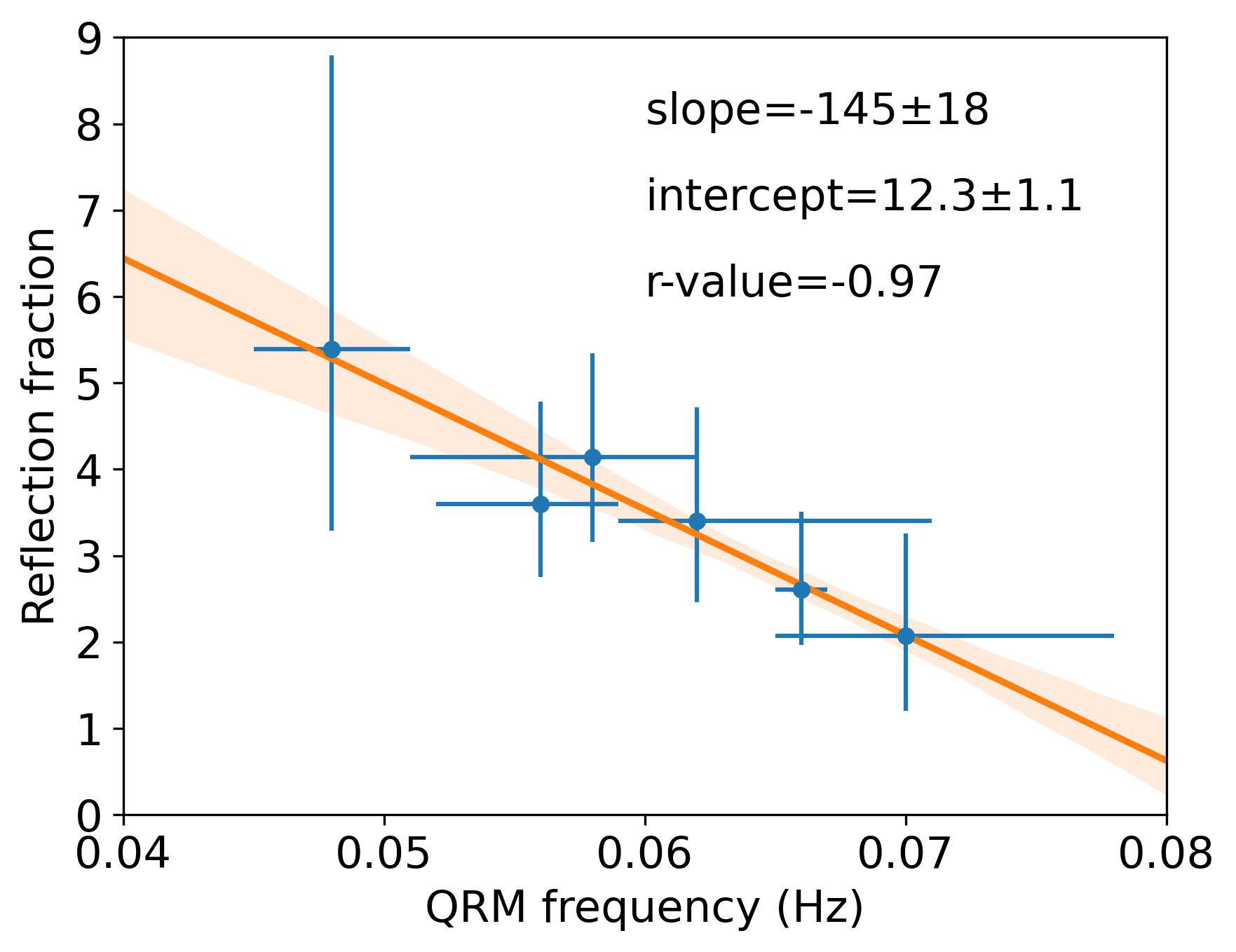}
   \caption{\textbf{Left:} Centroid frequency of QPOs versus reflection fraction and the correlation coefficient between them. \textbf{Right:} Centroid frequency of QRMs versus reflection fraction and the correlation coefficient between them. The centroid frequencies of both QPO and QRM are anti-correlated with the reflection fraction. The lines represent the best-fitting lines (90\% confidence interval) with a linear function of $\rm{y=slope*x+intercept}$, where y is the reflection fraction and x is the QPO/QRM frequency.}
   \label{figure4}
\end{figure*}

\begin{figure*}[!htbp]
   \centering
   \includegraphics[width=1.1\columnwidth]{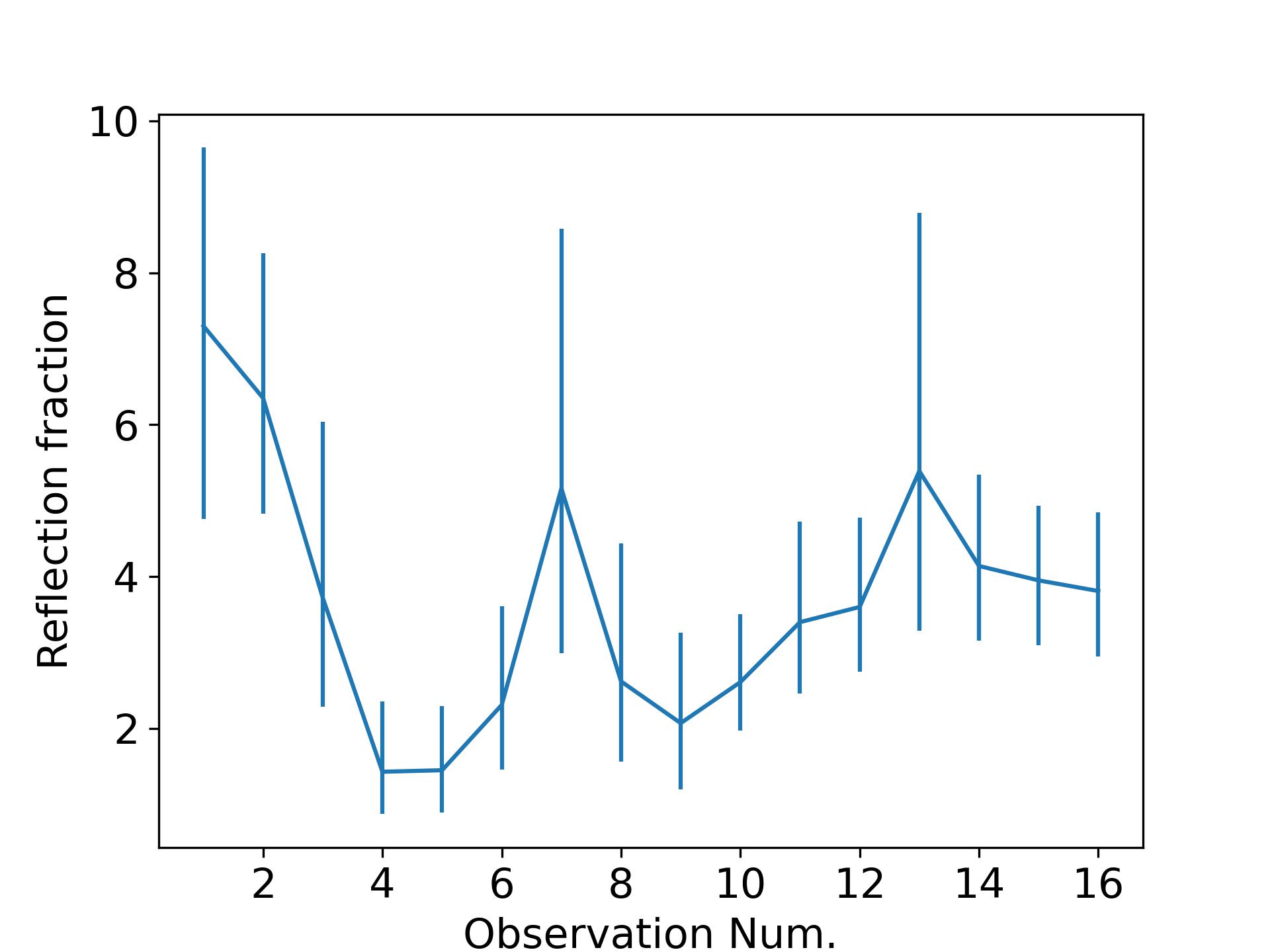}
   \includegraphics[width=\columnwidth]{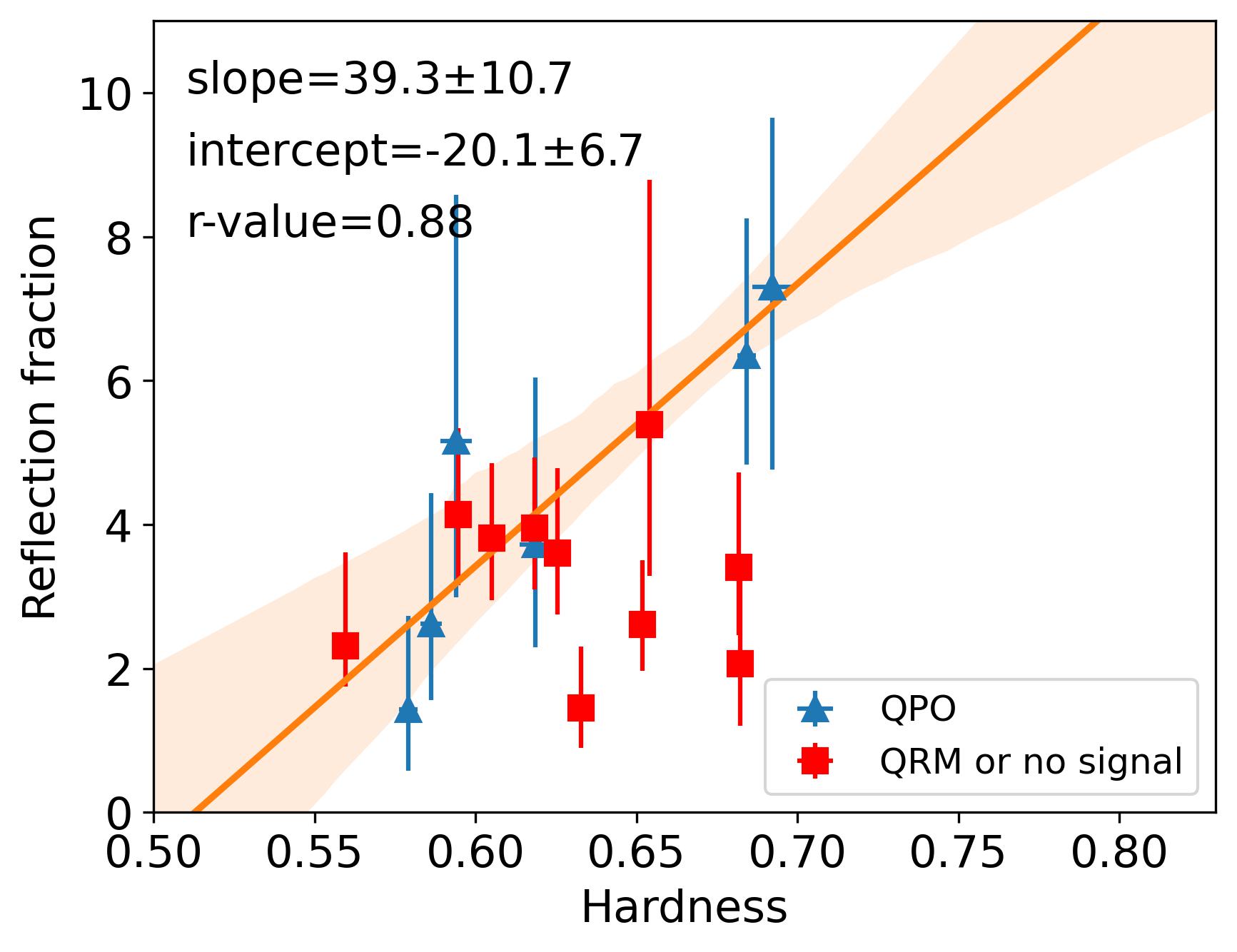}
   \caption{\textbf{Left:} Evolution of reflection fraction versus the observation time during the outburst. The reflection fraction declines in the first 8 observations when QPOs are detected and increases in the later observations when QRMs are detected.
   \textbf{Right:} Hardness ratio (between the LE 2-10 keV and ME 10-35 keV bands) versus reflection fraction for all the observations. Blue triangles are QPO observations, and red squares are QRM observations. The hardness ratio is positively correlated to the reflection fraction when the QPO signal is detected. The hardness ratio shows no relation to the reflection fraction in the case of QRMs or no signals. The line represents the best-fitting line (90\% confidence interval) with a liner function of $\rm{y=slope*x+intercept}$, where y is the reflection fraction and x is the hardness ratio.}
   \label{figure5}
\end{figure*}

The best-fitting spectral parameters are listed together in Table \ref{table2}. The results show that reflection features vary during the outburst. Figure \ref{figure3} presents the best-fitting spectra and the corresponding data-to-model ratios of observations Num. 4 and 10. Figure \ref{figure4} shows the relation of centroid frequencies of QPOs and QRMs versus reflection fraction, and the centroid frequencies of both QPOs and QRMs are anti-correlated to the reflection fraction with correlation coefficients of -0.97. The evolution of the reflection fraction is presented in the left panel of Figure \ref{figure5}. Comparing two figures (Figures \ref{figure1} and \ref{figure5}), we find that the reflection fraction has no relation to the fluxes of LE, ME, and HE bands. The hardness ratio positively correlates with the reflection fraction when QPOs are found (with a correlation coefficient of 0.88, see the right panel of Fig. \ref{figure5}). The data in Figures \ref{figure4} and \ref{figure5} are fitted with a linear function of $\rm{y=slope*x+intercept}$, where y is the reflection fraction and x is QPO/QRM frequency or hardness ratio. The lines in the figures represent the best-fitting lines with 90\% confidence intervals.

\section{discussion}


By comparing the QPOs/QRMs characteristics to the fitted reflection spectral parameters, we find that the QPO centroid frequency anti-correlates to the reflection fraction. The hardness ratio is positively correlated to the reflection fraction when QPOs are detected, and the inner radius of the accretion disk decreases with time when the hardness ratio declines. The decrease in the inner radius of the accretion disk can lead to a variation in the reflection fraction as well as the accretion rate \citep{2010MNRAS.409.1534D,2014MNRAS.444L.100D}. In addition, the precession of the inner accretion flow can lead to changes in reflection features \citep{2009MNRAS.397L.101I,2017MNRAS.464.2979I}. Therefore, the correlation between the hardness ratio and QPOs indicates that the QPOs may be generated by precession of the inner flow. The QRM centroid frequency also anti-correlates with the reflection fraction. However, the hardness ratio shows no relation to the reflection fraction when QRM appears. This indicates that QRM has a different physical origin from QPO. \citet{2022ApJ...937...33Y} calculated the mHz QRM energy-dependent fractional rms of 4U 1630-47 up to 100 keV. They showed that the mHz QRM phenomenon is more pronounced for non-thermal components, i.e., the corona region. Therefore, the QRM is probably generated from instabilities in the corona. These instabilities can lead to changes in both the corona structure and the reflection properties, and make the QRM centroid frequency and reflection fraction related. The appearance of mHz QRMs in 4U 1630-47 is related to the accretion rate, which can be interpreted by the local radiation-dominated disk instability model \citep{2008bhad.book.....K}. \citet{2022ApJ...937...33Y} suggested the mHz QRM could be caused by an unknown accretion instability arising from the corona, leading to flux modulation and variations in the reflection fraction.

To fit the spectra, observations 1-5 require a supersolar iron abundance. So far, no plausible physical explanation has been proffered for the black-hole systems to be iron-rich. Nonetheless, supersolar prediction for the Fe abundance has been reported in some stellar-mass black hole binaries such as GX 339-4 \citep{2015ApJ...813...84G}, V404 Cyg \citep{2017ApJ...839..110W}, and Cyg X-1 \citep{2015ApJ...808....9P}. A similar trend is found in AGNs as well, the iron of Seyfert galaxy 1H0707-495 is overabundant by a factor of 10-20 \citep{2009Natur.459..540F,2012MNRAS.422.1914D}. \citet{2018ASPC..515..282G} collected the reports of iron abundance obtained by reflection models for 13 AGNs and 9 BHBs, finding that iron abundance has a trend of a few times over the solar value in both AGNs and BHBs. Observed supersolar iron abundances are unlikely to be realistic, as metal enrichment mechanisms in AGNs and BHBs are expected to be different, and a possible explanation for the results is predicting a relatively low disk density \citep{2018ASPC..515..282G}. We set the disk density $n_e=10^{15}$cm$^{-3}$ in the fitting. However, the standard $\alpha$-disk model \citep{1973A&A....24..337S} and the results of 3D magneto-hydrodynamic (MHD) simulations \citep{2010ApJ...711..959N,2013ApJ...769..156S} suggest that the accretion disk of black holes has a density orders of magnitude larger than $n_e\sim 10^{15}$cm$^{-3}$. Therefore, we set the $\textrm{logN}$ to 20 (maximum of the model) and re-fit the data of Obs. 1. This still returns a supersolar iron abundance $A_{\textrm{Fe}}=2.3^{+3.4}_{-1.4}$, and the disk density may be higher than $10^{20}$cm$^{-3}$. Such a high disk density has been reported in MAXI J1836-194, \citet{2020MNRAS.493.2178D} reports that it has a high disk density $n_e\textgreater10^{21}$cm$^{-3}$. The iron abundance declines to a reasonable range in the later observations, which may be due to the increase in accretion rate and the decrease in the disk density. Since the fitting is insensitive to the value of disk density, the supersolar iron abundance problem could not be resolved with a maximum disk density of the model, we still set the disk density $n_e=10^{15}$cm$^{-3}$ in the fitting.

Our spectral fitting gives a low ionization of about $1.3-1.8$ and the ionization gradually increases along with the flux. \citet{2011ApJ...734..112B} used a standard $\alpha$-disk model \citep{1973A&A....24..337S} to derive an analytical estimate of $\xi(r)$, their estimation of $\xi$ is nicely independent of the central black hole mass. They also studied how the $\xi(r)$ should depend on various physical parameters such as black hole spin $a$, coronal dissipation fraction $f$ \citep{1994ApJ...436..599S}, and Eddington ratio ($\lambda = L_{\textrm{bol}} /L_{\textrm{Edd}}$, where $L_{\textrm{bol}}$ is the bolometric luminosity and $L_{\textrm{Edd}}$ is the Eddington luminosity). The value of $\xi$ is sensitive to Eddington ratio $\lambda$, spin $a$, the radius of peak reflection, and it can be very low at the inner disk for Eddington ratios of a few percent \citep[see Figure 2 in][]{2011ApJ...734..112B}. The Eddington ratio of the source is about 0.03 for Obs. 1 if assuming a black hole mass $M=10M_\sun$ and a distance of 4.7 kpc. Therefore, the low ionization is probably due to the low luminosity of the source and the high disk density.

A steep inner emissivity index ($\textgreater$ 8 in most observations) is needed to fit the spectra. Although the emissivity profile with index $q=3$ is considered as the standard, a steep emissivity index is commonly seen in X-ray binaries and AGNs. This has been reported in X-ray spectral analyses of AGNs, such as MCG-6-30-15 \citep{2002MNRAS.335L...1F,2004MNRAS.348.1415V,2007PASJ...59S.315M,2020MNRAS.498.3565T}, 1H0707-495 \citep{2009Natur.459..540F,2010MNRAS.401.2419Z,2011MNRAS.414.1269W,2012MNRAS.422.1914D}, and IRAS 13224-3809 \citep{2010MNRAS.406.2591P}, as well as in black hole binaries like XTE J1650-500 \citep{2004MNRAS.351..466M}, GX 339-4 \citep{2007ARA&A..45..441M}, Cyg X-1 \citep{2012MNRAS.424..217F} and GRS 1915+105 \citep{2020MNRAS.498.3565T}. The observed steep emissivity profiles can be attributed to the following scenarios: (1) a radial profile of the disk ionization or a compact centrally concentrated X-ray corona locates at a low height above the black hole, or the combination of the two \citep{2012A&A...545A.106S,2017MNRAS.472.1932G,2019MNRAS.487..550L}; (2) a ring-like or disk-like corona locates above the accretion disk with the axis parallel to the black hole \citep{2011MNRAS.414.1269W,2015MNRAS.449..129W}. Additionally, in the compact X-ray source, the radial profile of the disk irradiation shows a significant gradient, resulting in a significant decrease in the ionization parameter as the radius increases \citep{2019MNRAS.485..239K}. Neglecting this effect can lead to an overestimation of the radial emissivity index if the corona is highly compact and at a low height \citep{2012A&A...545A.106S}. The reflection model used in this work assumes the accretion disk has a constant ionization \citep{2014ApJ...782...76G,2014MNRAS.444L.100D}, which may also be the reason for obtaining a high radial emissivity index in the fitting.

\section{Conclusion}

In this work, we studied the quasi-periodic oscillations and spectral evolutions of 4U 1630-47 during its 2021 outburst based on \textit{Insight}-HXMT observations. Both quasi-periodic oscillations (QPOs) and quasi-regular modulations (QRMs) are detected in the outburst. The QPOs frequencies evolve from $\sim 1.6 - 3.6$ Hz, and QRMs have low frequencies around 0.05 - 0.07 Hz. The reflection components also evolve during the outburst. The QPOs centroid frequency is anti-correlated to the reflection fraction, which is consistent with the prediction of precessing inner flow model \citep{2009MNRAS.397L.101I} and provides evidence for a geometrical origin of QPOs. The centroid frequency of QRMs is also anti-correlated to the reflection fraction. The hardness ratio has a positive correlation with the reflection fraction when QPOs are detected, but shows no relation when the QRMs are detected. This hints that QRMs may have a different physical origin from QPOs. In addition, the mHz QRMs phenomenon is more pronounced for non-thermal emissions above 30 keV. Thus, we suggest that QRMs may be caused by instabilities in the corona.

\begin{table*}
\renewcommand{\arraystretch}{1.8}
    \centering
    \caption{Fitting results of \textit{Insight}-HXMT Observations of 4U 1630-47 in the 2021 outburst with model \texttt{constant*tbabs(diskbb+ relxillcp)}. The letter P indicates that the error of the parameter was pegged at the upper or lower boundary. All errors were calculated at 90 percent confidence level.}
    \begin{tabular}{c c c c c c c c c c c c}
       \hline
       Num. & Tbabs                     &  \multicolumn{2}{c}{diskbb} & \multicolumn{6}{c}{relxillcp} & $\chi^2_{\upsilon}$\\  \cline{2-10}
            & $N_{H}$                   &  ${T_{in}}$ & Norm  & ${R_{in}}$ & $q$ & $\Gamma$ & $\log\xi$ & ${A_{\textrm{Fe}}}$ & ${R_{f}}$ &  \\                     
            & $\times10^{22}$ cm$^{-3}$ &             &       &            &     &          &    erg cm s$^{-1}$       &            &           &                \\
       \hline
       1  & $6.2^{+0.6}_{-0.7}$  & $2.37^{+0.32}_{-0.22}$ & $3.2^{+3.0}_{-2.1}$ &   $-1.14^{+0.12}_{-0.13}$ & $6.8^{+2.4}_{-1.4}$ & $2.38^{+0.10}_{-0.12}$ & $1.3^{+0.6}_{-0.3}$ &                    $3.2^{+3.0}_{-2.1}$ & $7.3^{+2.4}_{-2.5}$ & 1.00 \\
       2  & $6.5^{+0.3}_{-0.3}$  & $2.18^{+0.14}_{-0.12}$ & $4.5^{+2.4}_{-1.7}$ &   $-1.26_{-0.04}^{0.07}$  & $10.0^{+P}_{-2.5}$     & $2.40^{+0.06}_{-0.04}$ & $1.4^{+0.3}_{-0.3}$ &                    $1.3^{+0.8}_{-0.5}$ & $6.4^{+1.9}_{-1.5}$ & 0.89 \\
       3  & $6.3^{+0.5}_{-0.7}$  & $2.62^{+0.40}_{-0.30}$ & $2.7^{+3.1}_{-1.5}$ &   $-1.00_{-0.25}^{+P}$     & $8.3^{+1.5}_{-1.8}$ & $2.29^{+0.07}_{-0.08}$ & $1.6^{+0.5}_{-0.5}$ &                    $3.3^{+2.9}_{-2.2}$ & $3.7^{+2.3}_{-1.4}$ & 0.99 \\
       4  & $6.9^{+0.3}_{-0.3}$  & $1.88^{+0.30}_{-0.27}$ & $5.6^{+5.1}_{-3.8}$ &   $-1.08_{-0.10}^{+0.07}$ & $8.4^{+1.4}_{-1.7}$ & $2.32^{+0.06}_{-0.07}$ & $1.7^{+1.0}_{-0.6}$ &                    $4.9^{+4.5}_{-3.3}$ & $1.4^{+0.9}_{-0.6}$ & 0.92 \\
       5  & $7.7^{+0.4}_{-0.5}$  & $1.90^{+0.45}_{-0.56}$ & $6.8^{+8.6}_{-5.8}$ &   $-1.12_{-0.17}^{+0.10}$ & $8.1^{+1.7}_{-2.4}$  & $2.40^{+0.06}_{-0.05}$ & $1.7^{+0.8}_{-0.6}$ &                    $1.9^{+3.0}_{-1.3}$ & $1.5^{+0.8}_{-0.6}$ & 1.02 \\
       6  & $7.0^{+0.4}_{-0.5}$  & $2.55^{+0.22}_{-0.19}$ & $6.1^{+4.0}_{-2.5}$ &  $-1.00_{-0.24}^{+P}$      & $10.0^{+P}_{-3.9}$      & $2.31^{+0.07}_{-0.07}$ & $1.6^{+0.5}_{-0.5}$ &                    $0.5^{+1.5}_{-P}$  & $2.3^{+1.3}_{-0.9}$ & 0.92 \\
       7  & $5.7^{+0.5}_{-0.5}$  & $2.60^{+0.35}_{-0.28}$ & $3.1^{+2.6}_{-1.6}$ &   $-1.16_{-0.10}^{+0.12}$ & $10.0^{+P}_{-4.0}$      & $2.39^{+0.11}_{-0.11}$ & $1.6^{+0.4}_{-0.5}$ &                    $0.5^{+2.1}_{-P}$     & $5.2^{+3.4}_{-2.2}$ & 0.97 \\
       8  & $6.4^{+0.4}_{-0.4}$  & $2.11^{+0.31}_{-0.18}$ & $5.0^{+4.6}_{-3.1}$ &   $-1.00_{-0.29}^{+P}$     & $6.2^{+3.0}_{-1.8}$ & $2.40^{+0.09}_{-0.09}$ & $1.5^{+0.4}_{-0.5}$ &                    $0.5^{+1.6}_{-P}$     & $2.6^{+1.8}_{-1.1}$ & 0.94 \\
       9  & $8.0^{+0.3}_{-0.3}$  & $2.58^{+0.22}_{-0.16}$ & $8.0^{+3.5}_{-3.0}$ &  $-1.00_{-0.12}^{+P}$      & $10.0^{+P}_{-2.8}$      & $2.37^{+0.06}_{-0.06}$ & $1.8^{+0.4}_{-0.5}$ &                    $0.5^{+2.8}_{-P}$     & $2.1^{+1.2}_{-0.9}$ & 0.98 \\
       10 & $8.1^{+0.3}_{-0.3}$  & $2.45^{+0.15}_{-0.12}$ & $11.1^{+3.9}_{-3.1}$ &  $-1.00_{-0.08}^{+P}$     & $10.0^{+P}_{-2.1}$      & $2.47^{+0.04}_{-0.03}$ & $1.7^{+0.4}_{-0.4}$ &                    $0.5^{+1.7}_{-P}$     & $2.6^{+0.9}_{-0.6}$ & 1.00 \\
       11 & $7.7^{+0.3}_{-0.3}$  & $2.36^{+0.13}_{-0.11}$ & $14.3^{+4.8}_{-3.8}$ &  $-1.05_{-0.06}^{+0.04}$ & $10.0^{+P}_{-2.4}$      & $2.39^{+0.05}_{-0.05}$ & $1.7^{+0.3}_{-0.5}$ &                    $0.5^{+1.4}_{-P}$     & $3.4^{+1.3}_{-0.9}$ & 1.00 \\
       12 & $7.7^{+0.3}_{-0.3}$  & $2.29^{+0.13}_{-0.12}$ & $16.5^{+6.0}_{-5.0}$ & $-1.00_{-0.07}^{+P}$      & $10.0^{+P}_{-1.7}$      & $2.43^{+0.04}_{-0.03}$ & $1.7^{+0.3}_{-0.3}$ &                    $0.5^{+1.4}_{-P}$     & $3.6^{+1.2}_{-0.9}$ & 0.94 \\
       13 & $8.0^{+0.5}_{-0.4}$  & $2.30^{+0.25}_{-0.16}$ & $14.8^{+8.4}_{-6.7}$ &  $-1.00_{-0.08}^{+P}$     & $10.0^{+P}_{-3.2}$      & $2.60^{+0.09}_{-0.07}$ & $1.6^{+0.4}_{-0.5}$ &                    $0.5^{+1.4}_{-P}$     & $5.4^{+3.4}_{-2.1}$ & 0.96 \\
       14 & $7.5^{+0.3}_{-0.2}$  & $2.35^{+0.06}_{-0.06}$ & $22.6^{+4.0}_{-3.6}$ &  $-1.00^{+P}_{-0.02}$     & $10.0^{+P}_{-1.7}$      & $2.44^{+0.04}_{-0.03}$ & $1.6^{+0.2}_{-0.3}$ &                    $0.5^{+0.3}_{-P}$     & $4.1^{+1.2}_{-1.0}$ & 0.98 \\
       15 & $7.8^{+0.2}_{-0.2}$  & $2.41^{+0.07}_{-0.05}$ & $23.3^{+3.2}_{-3.8}$ &  $-1.00^{+P}_{-0.02}$     & $10.0^{+P}_{-1.5}$      & $2.46^{+0.04}_{-0.03}$ & $1.7^{+0.2}_{-0.2}$ &                    $0.5^{+0.2}_{-P}$     & $4.0^{+1.0}_{-0.9}$ & 0.95 \\
       16 & $7.7^{+0.3}_{-0.2}$  & $2.39^{+0.06}_{-0.05}$ & $26.4^{+3.8}_{-4.2}$ &  $-1.00^{+P}_{-0.02}$     & $10.0^{+P}_{-1.6}$      & $2.43^{+0.04}_{-0.03}$ & $1.8^{+0.3}_{-0.3}$ &                    $0.5^{+0.3}_{-P}$     & $3.8^{+1.0}_{-0.9}$ & 1.00 \\
       \hline
    \end{tabular}
    \label{table2}
\end{table*}

\section*{Acknowledgements}
We are grateful to the referee for the fruitful comments to improve the manuscript. This work is supported by the National Key Research and Development Program of China (Grants No. 2021YFA0718503 and 2023YFA1607901), the NSFC (12133007). This work has made use of data from the \textit{Insight-}HXMT mission, a project funded by the China National Space Administration (CNSA) and the Chinese Academy of Sciences (CAS).

\bibliography{bib_4U_1630-47}{}

\begin{thebibliography}{}
\expandafter\ifx\csname natexlab\endcsname\relax\def\natexlab#1{#1}\fi
\providecommand{\url}[1]{\href{#1}{#1}}
\providecommand{\dodoi}[1]{doi:~\href{http://doi.org/#1}{\nolinkurl{#1}}}
\providecommand{\doeprint}[1]{\href{http://ascl.net/#1}{\nolinkurl{http://ascl.net/#1}}}
\providecommand{\doarXiv}[1]{\href{https://arxiv.org/abs/#1}{\nolinkurl{https://arxiv.org/abs/#1}}}

\bibitem[{Altamirano \& Strohmayer(2012)}]{2012ApJ...754L..23A}
Altamirano, D., \& Strohmayer, T. 2012, \href{https://ui.adsabs.harvard.edu/abs/2012ApJ...754L..23A/abstract}{\apj}, 751, 23

\bibitem[{Altamirano {et~al.}(2011)Altamirano, Belloni, Linares, van~der Klis, Wijnands, Curran, Kalamkar, Stiele, Motta, Muñoz-Darias, Casella, \& Krimm}]{2011ApJ...742L..17A}
Altamirano, D., Belloni, T., Linares, M., {et~al.} 2011, \href{https://ui.adsabs.harvard.edu/abs/2011ApJ...742L..17A/abstract}{\apj}, 742, L17

\bibitem[{Arnaud(1996)}]{1996ASPC..101...17A}
Arnaud, K.~A. 1996, \href{https://ui.adsabs.harvard.edu/abs/1996ASPC..101...17A/abstract}{ASPC}, 101, 17

\bibitem[{Ballantyne {et~al.}(2011)Ballantyne, McDuffie, \& Rusin}]{2011ApJ...734..112B}
Ballantyne, D.~R., McDuffie, J.~R., \& Rusin, J.~S. 2011, \href{https://ui.adsabs.harvard.edu/abs/2011ApJ...734..112B/abstract}{\apj}, 734, 112

\bibitem[{Belloni \& Hasinger(1990)}]{1990A&A...230..103B}
Belloni, T., \& Hasinger, G. 1990, \href{https://ui.adsabs.harvard.edu/abs/1990A&A...230..103B/abstract}{A\&A}, 230, 103

\bibitem[{Belloni {et~al.}(2005)Belloni, Homan, Casella, van~der Klis, Nespoli, Lewin, Miller, \& Méndez}]{2005A&A...440..207B}
Belloni, T., Homan, J., Casella, P., {et~al.} 2005, \href{https://ui.adsabs.harvard.edu/abs/2005A&A...440..207B/abstract}{\aa}, 440, 207

\bibitem[{Belloni {et~al.}(2006)Belloni, Soleri, Casella, Méndez, \& Migliari}]{2006MNRAS.369..305B}
Belloni, T., Soleri, P., Casella, P., Méndez, M., \& Migliari, S. 2006, \href{https://ui.adsabs.harvard.edu/abs/2006MNRAS.369..305B/abstract}{\mnras}, 369, 305

\bibitem[{Belloni(2010)}]{2010LNP...794...53B}
Belloni, T.~M. 2010, Lecture Notes in Physics, Vol. 794, \href{https://ui.adsabs.harvard.edu/abs/2010LNP...794...53B/abstract}{States and Transitions in Black Hole Binaries} (Springer-Verlag)

\bibitem[{Belloni \& Motta(2016)}]{2016ASSL..440...61B}
Belloni, T.~M., \& Motta, S.~E. 2016, \href{https://ui.adsabs.harvard.edu/abs/2016ASSL..440...61B/abstract}{ASSL}, 440, 61

\bibitem[{Bu {et~al.}(2015)Bu, Chen, Li, Qu, Belloni, \& Zhang}]{2015ApJ...799....2B}
Bu, Q.-c., Chen, L., Li, Z.-s., {et~al.} 2015, \href{https://ui.adsabs.harvard.edu/abs/2015ApJ...799....2B/abstract}{\apj}, 799, 2

\bibitem[{Cao {et~al.}(2020)Cao, Jiang, Meng, Zhang, Luo, Yang, Zhang, Gu, Sun, Liu, {et~al.}}]{cao2020medium}
Cao, X., Jiang, W., Meng, B., {et~al.} 2020, SCIENCE CHINA Physics, Mechanics \& Astronomy, 63, 1

\bibitem[{Casella {et~al.}(2005)Casella, Belloni, \& Stella}]{2005ApJ...629..403C}
Casella, P., Belloni, T., \& Stella, L. 2005, \href{https://ui.adsabs.harvard.edu/abs/2005ApJ...629..403C/abstract}{\apj}, 629, 403

\bibitem[{Chakrabarti {et~al.}(2008)Chakrabarti, Debnath, Nandi, \& Pal}]{2008A&A...489L..41C}
Chakrabarti, S.~K., Debnath, D., Nandi, A., \& Pal, P.~S. 2008, \href{https://ui.adsabs.harvard.edu/abs/2008A&A...489L..41C/abstract}{A\&A}, 489, L41

\bibitem[{Chen {et~al.}(1997)Chen, Shrader, \& Livio}]{1997ApJ...491..312C}
Chen, W., Shrader, C.~R., \& Livio, M. 1997, \href{https://ui.adsabs.harvard.edu/abs/1997ApJ...491..312C/abstract}{ApJ}, 491, 312

\bibitem[{Chen {et~al.}(2021)Chen, Wang, Tang, Ding, Tuo, Mushtukov, Nishimura, Zhang, Ge, Song, {et~al.}}]{chen2021relation}
Chen, X., Wang, W., Tang, Y., {et~al.} 2021, The Astrophysical Journal, 919, 33

\bibitem[{Chen {et~al.}(2020)Chen, Cui, Li, Wang, Xu, Lu, Wang, Chen, Han, Hu, {et~al.}}]{chen2020low}
Chen, Y., Cui, W., Li, W., {et~al.} 2020, Science China Physics, Mechanics \& Astronomy, 63, 1

\bibitem[{Dauser {et~al.}(2014)Dauser, Garcia, Parker, Fabian, \& Wilms}]{2014MNRAS.444L.100D}
Dauser, T., Garcia, J., Parker, M.~L., Fabian, A.~C., \& Wilms, J. 2014, \href{https://ui.adsabs.harvard.edu/abs/2014MNRAS.444L.100D/abstract}{\mnras}, 444, L100

\bibitem[{Dauser {et~al.}(2013)Dauser, Garcia, Wilms, Böck, Brenneman, Falanga, Fukumura, \& Reynolds}]{2013MNRAS.430.1694D}
Dauser, T., Garcia, J., Wilms, J., {et~al.} 2013, \href{https://ui.adsabs.harvard.edu/abs/2013MNRAS.430.1694D/abstract}{MNRAS}, 430, 1694

\bibitem[{Dauser {et~al.}(2016)Dauser, García, Walton, Eikmann, Kallman, McClintock, \& Wilms}]{2016A&A...590A..76D}
Dauser, T., García, J., Walton, D.~J., {et~al.} 2016, \href{https://ui.adsabs.harvard.edu/abs/2016A&A...590A..76D/abstract}{A\&A}, 290, A76

\bibitem[{Dauser {et~al.}(2010)Dauser, Wilms, Reynolds, \& Brenneman}]{2010MNRAS.409.1534D}
Dauser, T., Wilms, J., Reynolds, C.~S., \& Brenneman, L.~W. 2010, \href{https://ui.adsabs.harvard.edu/abs/2010MNRAS.409.1534D/abstract}{\mnras}, 409, 1534

\bibitem[{Dauser {et~al.}(2012)Dauser, Svoboda, Schartel, Wilms, Dovčiak, Ehle, Karas, Santos-Lleó, \& Marshall}]{2012MNRAS.422.1914D}
Dauser, T.~s., Svoboda, J.~s., Schartel, N.~s., {et~al.} 2012, \href{https://ui.adsabs.harvard.edu/abs/2012MNRAS.422.1914D/abstract}{\mnras}, 422, 1914

\bibitem[{Dong {et~al.}(2020)Dong, García, Liu, Zhao, Zheng, \& Gou}]{2020MNRAS.493.2178D}
Dong, Y., García, J.~A., Liu, Z., {et~al.} 2020, \href{https://ui.adsabs.harvard.edu/abs/2020MNRAS.493.2178D/abstract}{\mnras}, 493, 2178

\bibitem[{Eardley {et~al.}(1975)Eardley, Lightman, \& Shapiro}]{1975ApJ...199L.153E}
Eardley, D.~M., Lightman, A.~P., \& Shapiro, S.~L. 1975, \href{https://ui.adsabs.harvard.edu/abs/1975ApJ...199L.153E/abstract}{\apj}, 199, 153

\bibitem[{Fabian {et~al.}(1989)Fabian, Rees, Stella, \& White}]{1989MNRAS.238..729F}
Fabian, A.~C., Rees, M.~J., Stella, L., \& White, N.~E. 1989, \href{https://ui.adsabs.harvard.edu/abs/1989MNRAS.238..729F/abstract}{\mnras}, 238, 729

\bibitem[{Fabian {et~al.}(2002)Fabian, Vaughan, Nandra, Iwasawa, Ballantyne, Lee, De~Rosa, Turner, \& Young}]{2002MNRAS.335L...1F}
Fabian, A.~C., Vaughan, S., Nandra, K., {et~al.} 2002, \href{https://ui.adsabs.harvard.edu/abs/2002MNRAS.335L...1F/abstract}{\mnras}, 335, L1

\bibitem[{Fabian {et~al.}(2009)Fabian, Zoghbi, Ross, Uttley, Gallo, Brandt, Blustin, Boller, Caballero-Garcia, Larsson, Miller, Miniutti, Ponti, Reis, Reynolds, Tanaka, \& Young}]{2009Natur.459..540F}
Fabian, A.~C., Zoghbi, A., Ross, R.~R., {et~al.} 2009, \href{https://ui.adsabs.harvard.edu/abs/2009Natur.459..540F/abstract}{Natur.}, 459, 540

\bibitem[{Fabian {et~al.}(2012)Fabian, Wilkins, Miller, Reis, Reynolds, Cackett, Nowak, Pooley, Pottschmidt, Sanders, Ross, \& Wilms}]{2012MNRAS.424..217F}
Fabian, A.~C., Wilkins, D.~R., Miller, J.~M., {et~al.} 2012, \href{https://ui.adsabs.harvard.edu/abs/2012MNRAS.424..217F/abstract}{\mnras}, 424, 217

\bibitem[{García \& Kallman(2010)}]{2010ApJ...718..695G}
García, J., \& Kallman, T.~R. 2010, \href{https://ui.adsabs.harvard.edu/abs/2010ApJ...718..695G/abstract}{\apj}, 719, 695

\bibitem[{García {et~al.}(2014)García, Dauser, Lohfink, Kallman, Steiner, McClintock, Brenneman, Wilms, Eikmann, Reynolds, \& Tombesi}]{2014ApJ...782...76G}
García, J., Dauser, T., Lohfink, A., {et~al.} 2014, \href{https://ui.adsabs.harvard.edu/abs/2014ApJ...782...76G/abstract}{\apj}, 782, 76

\bibitem[{García {et~al.}(2018)García, Kallman, Bautista, Mendoza, Deprince, Palmeri, \& Quinet}]{2018ASPC..515..282G}
García, J.~A., Kallman, T.~R., Bautista, M., {et~al.} 2018, \href{https://ui.adsabs.harvard.edu/abs/2018ASPC..515..282G/abstract}{ASPC}, 515, 282

\bibitem[{García {et~al.}(2015)García, Steiner, McClintock, Remillard, Grinberg, \& Dauser}]{2015ApJ...813...84G}
García, J.~A., Steiner, J.~F., McClintock, J.~E., {et~al.} 2015, \href{https://ui.adsabs.harvard.edu/abs/2015ApJ...813...84G/abstract}{\apj}, 813, 84

\bibitem[{Gonzalez {et~al.}(2017)Gonzalez, Wilkins, \& Gallo}]{2017MNRAS.472.1932G}
Gonzalez, A.~G., Wilkins, D.~R., \& Gallo, L.~C. 2017, \href{https://ui.adsabs.harvard.edu/abs/2017MNRAS.472.1932G/abstract}{\mnras}, 472, 1932

\bibitem[{Grindlay {et~al.}(2014)Grindlay, Miller, \& Tang}]{2014AAS...22340606G}
Grindlay, J.~E., Miller, G.~F., \& Tang, S. 2014, \href{https://ui.adsabs.harvard.edu/abs/2014AAS...22340606G/abstract}{AAS}, 223, 406.06

\bibitem[{Heil {et~al.}(2015)Heil, Uttley, \& Klein-Wolt}]{2015MNRAS.448.3348H}
Heil, L.~M., Uttley, P., \& Klein-Wolt, M. 2015, \href{https://ui.adsabs.harvard.edu/abs/2015MNRAS.448.3348H/abstract}{\mnras}, 448, 3348

\bibitem[{Homan {et~al.}(2005)Homan, Buxton, Markoff, Bailyn, Nespoli, \& Belloni}]{2005ApJ...624..295H}
Homan, J., Buxton, M., Markoff, S., {et~al.} 2005, \href{https://ui.adsabs.harvard.edu/abs/2005ApJ...624..295H/abstract}{\apj}, 624, 295

\bibitem[{Homan {et~al.}(2003)Homan, Klein-Wolt, Rossi, Miller, Wijnands, Belloni, van~der Klis, \& Lewin}]{2003ApJ...586.1262H}
Homan, J., Klein-Wolt, M., Rossi, S., {et~al.} 2003, \href{https://ui.adsabs.harvard.edu/abs/2003ApJ...586.1262H/abstract}{\apj}, 586, 1262

\bibitem[{Homan {et~al.}(2001)Homan, Wijnands, van~der Klis, Belloni, van Paradijs, Klein-Wolt, Fender, \& Méndez}]{2001ApJS..132..377H}
Homan, J., Wijnands, R., van~der Klis, M., {et~al.} 2001, \href{https://ui.adsabs.harvard.edu/abs/2001ApJS..132..377H/abstract}{ApJS}, 132, 377

\bibitem[{Ichimaru(1977)}]{1977ApJ...214..840I}
Ichimaru, S. 1977, \href{https://ui.adsabs.harvard.edu/abs/1977ApJ...214..840I/abstract}{\apj}, 214, 840

\bibitem[{Ingram {et~al.}(2009)Ingram, Done, \& Fragile}]{2009MNRAS.397L.101I}
Ingram, A., Done, C., \& Fragile, P.~C. 2009, \href{https://ui.adsabs.harvard.edu/abs/2009MNRAS.397L.101I/abstract}{\mnras}, 397, L101

\bibitem[{Ingram {et~al.}(2017)Ingram, van~der Klis, Middleton, Altamirano, \& Uttley}]{2017MNRAS.464.2979I}
Ingram, A., van~der Klis, M., Middleton, M., Altamirano, D., \& Uttley, P. 2017, \href{https://ui.adsabs.harvard.edu/abs/2017MNRAS.464.2979I/abstract}{\mnras}, 464, 2979

\bibitem[{Ingram {et~al.}(2016)Ingram, van~der Klis, Middleton, Done, Altamirano, Heil, Uttley, \& Axelsson}]{2016MNRAS.461.1967I}
Ingram, A., van~der Klis, M., Middleton, M., {et~al.} 2016, \href{https://ui.adsabs.harvard.edu/abs/2016MNRAS.461.1967I/abstract}{\mnras}, 461, 1967

\bibitem[{Ingram \& Motta(2019)}]{2019NewAR..8501524I}
Ingram, A.~R., \& Motta, S.~E. 2019, \href{https://ui.adsabs.harvard.edu/abs/2019NewAR..8501524I/abstract}{New Astron. Rev.}, 85, id. 101524

\bibitem[{Jones {et~al.}(1976)Jones, Forman, Tananbaum, \& Turner}]{1976ApJ...210L...9J}
Jones, C., Forman, W., Tananbaum, H., \& Turner, M. J.~L. 1976, \href{https://ui.adsabs.harvard.edu/abs/1976ApJ...210L...9J/abstract}{\apj}, 210, L9

\bibitem[{Kalemci \& Maccarone(2018)}]{2018ApJ...859...88K}
Kalemci, E., \& Maccarone, T. J.and~Tomsick, J.~A. 2018, \href{https://ui.adsabs.harvard.edu/abs/2018ApJ...859...88K/abstract}{\apj}, 859, 88

\bibitem[{Kammoun {et~al.}(2019)Kammoun, Domček, Svoboda, Dovčiak, \& Matt}]{2019MNRAS.485..239K}
Kammoun, E.~S., Domček, V., Svoboda, J., Dovčiak, M., \& Matt, G. 2019, \href{https://ui.adsabs.harvard.edu/abs/2019MNRAS.485..239K/abstract}{\mnras}, 485, 239

\bibitem[{Kato(2001)}]{2001PASJ...53....1K}
Kato, S. 2001, \href{https://ui.adsabs.harvard.edu/abs/2001PASJ...53....1K/abstract}{PASJ}, 53, 1

\bibitem[{Kato \& Fukue(1980)}]{1980PASJ...32..377K}
Kato, S., \& Fukue, J. 1980, \href{https://ui.adsabs.harvard.edu/abs/1980PASJ...32..377K/abstract}{PASJ}, 32, 377

\bibitem[{Kato {et~al.}(2008)Kato, Fukue, \& Mineshige}]{2008bhad.book.....K}
Kato, S., Fukue, J., \& Mineshige, S. 2008, \href{https://ui.adsabs.harvard.edu/abs/2008bhad.book.....K/abstract}{Black-Hole Accretion Disks --- Towards a New Paradigm --- } (Kyoto University Press)

\bibitem[{King {et~al.}(2014)King, Walton, Miller, Barret, Boggs, Christensen, Craig, Fabian, Fürst, Hailey, Harrison, Krivonos, Mori, Natalucci, Stern, Tomsick, \& Zhang}]{2014ApJ...784L...2K}
King, A.~L., Walton, D.~J., Miller, J.~M., {et~al.} 2014, \href{https://ui.adsabs.harvard.edu/abs/2014ApJ...784L...2K/abstract}{\apj}, 784, L2

\bibitem[{Li {et~al.}(2020)Li, Li, Tan, Yang, Ge, Zhang, Tuo, Wu, Liao, Zhang, Song, Zhang, Qu, Zhang, Lu, Xu, Liu, Cao, Chen, \& Nie}]{2020JHEAp..27...64L}
Li, X., Li, X., Tan, Y., {et~al.} 2020, \href{https://ui.adsabs.harvard.edu/abs/2020JHEAp..27...64L/abstract}{JHEAp}, 27, 64

\bibitem[{Lightman \& Eardley(1974)}]{1974ApJ...187L...1L}
Lightman, A.~P., \& Eardley, D.~M. 1974, \href{https://ui.adsabs.harvard.edu/abs/1974ApJ...187L...1L/abstract}{\apj}, 187, L2

\bibitem[{Lightman \& Rybicki(1980)}]{1980ApJ...236..928L}
Lightman, A.~P., \& Rybicki, G.~B. 1980, \href{https://ui.adsabs.harvard.edu/abs/1980ApJ...236..928L/abstract}{\apj}, 236, 928

\bibitem[{Liska {et~al.}(2019)Liska, Tchekhovskoy, Ingram, \& van~der Klis}]{2019MNRAS.487..550L}
Liska, M., Tchekhovskoy, A., Ingram, A., \& van~der Klis, M. 2019, \href{https://ui.adsabs.harvard.edu/abs/2019MNRAS.487..550L/abstract}{\mnras}, 487, 550

\bibitem[{Liu {et~al.}(2020)Liu, Zhang, Li, Lu, Chang, Li, Zhang, Jin, Yu, Zhang, {et~al.}}]{liu2020high}
Liu, C., Zhang, Y., Li, X., {et~al.} 2020, SCIENCE CHINA Physics, Mechanics \& Astronomy, 63, 1

\bibitem[{Liu {et~al.}(2022)Liu, Liu, Bambi, \& Ji}]{2022MNRAS.512.2082L}
Liu, Q., Liu, H., Bambi, C., \& Ji, L. 2022, \href{https://ui.adsabs.harvard.edu/abs/2022MNRAS.512.2082L/abstract}{\mnras}, 512, 2082

\bibitem[{Miller(2007)}]{2007ARA&A..45..441M}
Miller, J.~M. 2007, \href{https://ui.adsabs.harvard.edu/abs/2007ARA&A..45..441M/abstract}{ARA\&A}, 45, 441

\bibitem[{Miniutti {et~al.}(2004)Miniutti, Fabian, \& Miller}]{2004MNRAS.351..466M}
Miniutti, G., Fabian, A.~C., \& Miller, J.~M. 2004, \href{https://ui.adsabs.harvard.edu/abs/2004MNRAS.351..466M/abstract}{\mnras}, 351, 466

\bibitem[{Miyamoto {et~al.}(1991)Miyamoto, Kimura, Kitamoto, Dotani, \& Ebisawa}]{1991ApJ...383..784M}
Miyamoto, S., Kimura, K., Kitamoto, S., Dotani, T., \& Ebisawa, K. 1991, \href{https://ui.adsabs.harvard.edu/abs/1991ApJ...383..784M/abstract}{\apj}, 383, 784

\bibitem[{Molteni {et~al.}(1996)Molteni, Sponholz, \& Chakrabarti}]{1996ApJ...457..805M}
Molteni, D., Sponholz, H., \& Chakrabarti, S.~K. 1996, \href{https://ui.adsabs.harvard.edu/abs/1996ApJ...457..805M/abstract}{\apj}, 457, 805

\bibitem[{Morgan {et~al.}(1997)Morgan, Remillard, \& Greiner}]{1997ApJ...482..993M}
Morgan, E.~H., Remillard, R.~A., \& Greiner, J. 1997, \href{https://ui.adsabs.harvard.edu/abs/1997ApJ...482..993M/abstract}{\apj}, 482, 993

\bibitem[{Motta {et~al.}(2012)Motta, Homan, Muñoz~Darias, Casella, Belloni, Hiemstra, \& Méndez}]{2012MNRAS.427..595M}
Motta, S., Homan, J., Muñoz~Darias, T., {et~al.} 2012, \href{https://ui.adsabs.harvard.edu/abs/2012MNRAS.427..595M/abstract}{\mnras}, 427, 595

\bibitem[{Motta {et~al.}(2011)Motta, Muñoz-Darias, Casella, Belloni, \& Homan}]{2011MNRAS.418.2292M}
Motta, S., Muñoz-Darias, T., Casella, P., Belloni, T., \& Homan, J. 2011, \href{https://ui.adsabs.harvard.edu/abs/2011MNRAS.418.2292M/abstract}{\mnras}, 418, 2292

\bibitem[{Motta {et~al.}(2015)Motta, Casella, Henze, Muñoz-Darias, Sanna, Fender, \& Belloni}]{2015MNRAS.447.2059M}
Motta, S.~E., Casella, P., Henze, M., {et~al.} 2015, \href{https://ui.adsabs.harvard.edu/abs/2015MNRAS.447.2059M/abstract}{\mnras}, 447, 2059

\bibitem[{Muñoz-Darias {et~al.}(2011)Muñoz-Darias, Motta, \& Belloni}]{2011MNRAS.410..679M}
Muñoz-Darias, T., Motta, S., \& Belloni, T.~M. 2011, \href{https://ui.adsabs.harvard.edu/abs/2011MNRAS.410..679M/abstract}{\mnras}, 410, 679

\bibitem[{Neilsen {et~al.}(2012)Neilsen, Remillard, \& Lee}]{2012ApJ...750...71N}
Neilsen, J., Remillard, R.~A., \& Lee, J.~C. 2012, \href{https://ui.adsabs.harvard.edu/abs/2012ApJ...750...71N/abstract}{\apj}, 750, 71

\bibitem[{Noble {et~al.}(2010)Noble, Krolik, \& Hawley}]{2010ApJ...711..959N}
Noble, S.~C., Krolik, J.~H., \& Hawley, J.~F. 2010, \href{https://ui.adsabs.harvard.edu/abs/2010ApJ...711..959N/abstract}{\apj}, 711, 959

\bibitem[{Novikov \& Thorne(1973)}]{1973blho.conf..343N}
Novikov, I.~D., \& Thorne, K.~S. 1973, \href{https://ui.adsabs.harvard.edu/abs/1973blho.conf..343N/abstract}{Black holes (Les astres occlus)}, -, 343

\bibitem[{Parker {et~al.}(2015)Parker, Tomsick, Miller, Yamaoka, Lohfink, Nowak, Fabian, Alston, Boggs, Christensen, Craig, Fürst, Gandhi, Grefenstette, Grinberg, Hailey, Harrison, Kara, King, \& Stern}]{2015ApJ...808....9P}
Parker, M.~L., Tomsick, J.~A., Miller, J.~M., {et~al.} 2015, \href{https://ui.adsabs.harvard.edu/abs/2015ApJ...808....9P/abstract}{\apj}, 808, 9

\bibitem[{Parmar {et~al.}(1995)Parmar, Angelini, \& White}]{1995ApJ...452L.129P}
Parmar, A.~N., Angelini, L., \& White, N.~E. 1995, \href{https://ui.adsabs.harvard.edu/abs/1995ApJ...452L.129P/abstract}{\apj}, 452, L129

\bibitem[{Ponti {et~al.}(2010)Ponti, Gallo, Fabian, Miniutti, Zoghbi, Uttley, Ross, Vasudevan, Tanaka, \& Brandt}]{2010MNRAS.406.2591P}
Ponti, G., Gallo, L.~C., Fabian, A.~C., {et~al.} 2010, \href{https://ui.adsabs.harvard.edu/abs/2010MNRAS.406.2591P/abstract}{\mnras}, 406, 2591

\bibitem[{Priedhorsky(1986)}]{1986Ap&SS.126...89P}
Priedhorsky, W. 1986, \href{https://ui.adsabs.harvard.edu/abs/1986Ap&SS.126...89P/abstract}{Ap\&SS}, 126, 89

\bibitem[{Remillard \& McClintock(2006)}]{2006ARA&A..44...49R}
Remillard, R.~A., \& McClintock, J.~E. 2006, \href{https://ui.adsabs.harvard.edu/abs/2006ARA&A..44...49R/abstract}{ARA\&A}, 44, 49

\bibitem[{Remillard {et~al.}(1999)Remillard, Morgan, McClintock, Bailyn, \& Orosz}]{1999ApJ...522..397R}
Remillard, R.~A., Morgan, E.~H., McClintock, J.~E., Bailyn, C.~D., \& Orosz, J.~A. 1999, \href{https://ui.adsabs.harvard.edu/abs/1999ApJ...522..397R/abstract}{\apj}, 522, 397

\bibitem[{Remillard {et~al.}(2002)Remillard, Muno, McClintock, \& Orosz}]{2002ApJ...580.1030R}
Remillard, R.~A., Muno, M.~P., McClintock, J.~E., \& Orosz, J.~A. 2002, \href{https://ui.adsabs.harvard.edu/abs/2002ApJ...580.1030R/abstract}{\apj}, 580, 1030

\bibitem[{Reynolds \& Begelman(1997)}]{1997ApJ...488..109R}
Reynolds, C.~S., \& Begelman, M.~C. 1997, \href{https://ui.adsabs.harvard.edu/abs/1997ApJ...488..109R/abstract}{\apj}, 488, 109

\bibitem[{Russell {et~al.}(2020)Russell, Casella, Kalemci, Vahdat~Motlagh, Saikia, Pirbhoy, \& Maitra}]{2020MNRAS.495..182R}
Russell, D.~M., Casella, P., Kalemci, E., {et~al.} 2020, \href{https://ui.adsabs.harvard.edu/abs/2020MNRAS.495..182R/abstract}{\mnras}, 495, 182

\bibitem[{Schnittman {et~al.}(2006)Schnittman, Homan, \& Miller}]{2006ApJ...642..420S}
Schnittman, J.~D., Homan, J., \& Miller, J.~M. 2006, \href{https://ui.adsabs.harvard.edu/abs/2006ApJ...642..420S/abstract}{\apj}, 642, 420

\bibitem[{Schnittman {et~al.}(2013)Schnittman, Krolik, \& Noble}]{2013ApJ...769..156S}
Schnittman, J.~D., Krolik, J.~H., \& Noble, S.~C. 2013, \href{https://ui.adsabs.harvard.edu/abs/2013ApJ...769..156S/abstract}{\apj}, 769, 156

\bibitem[{Seifina {et~al.}(2014)Seifina, Titarchuk, \& Shaposhnikov}]{2014ApJ...789...57S}
Seifina, E., Titarchuk, L., \& Shaposhnikov, N. 2014, \href{https://ui.adsabs.harvard.edu/abs/2014ApJ...789...57S/abstract}{\apj}, 789, 57

\bibitem[{Shakura \& Sunyaev(1973)}]{1973A&A....24..337S}
Shakura, N.~I., \& Sunyaev, R.~A. 1973, \href{https://ui.adsabs.harvard.edu/abs/1973A&A....24..337S/abstract}{A\&A}, 24, 337

\bibitem[{Sriram {et~al.}(2012)Sriram, Rao, \& Choi}]{2012A&A...541A...6S}
Sriram, K., Rao, A.~R., \& Choi, C.~S. 2012, \href{https://ui.adsabs.harvard.edu/abs/2012A&A...541A...6S/abstract}{\aa}, 541, 6

\bibitem[{Sriram {et~al.}(2013)Sriram, Rao, \& Choi}]{2013ApJ...775...28S}
---. 2013, \href{https://ui.adsabs.harvard.edu/abs/2013ApJ...775...28S/abstract}{\apj}, 775, 28

\bibitem[{Stella \& Vietri(1998)}]{1998ApJ...492L..59S}
Stella, L., \& Vietri, M. 1998, \href{https://ui.adsabs.harvard.edu/abs/1998ApJ...492L..59S/abstract}{\apj}, 492, L59

\bibitem[{Stella {et~al.}(1999)Stella, Vietri, \& Morsink}]{1999ApJ...524L..63S}
Stella, L., Vietri, M., \& Morsink, S.~M. 1999, \href{https://ui.adsabs.harvard.edu/abs/1999ApJ...524L..63S/abstract}{\apj}, 524, L63

\bibitem[{Strohmayer(2001)}]{2001ApJ...554L.169S}
Strohmayer, T.~E. 2001, \href{https://ui.adsabs.harvard.edu/abs/2001ApJ...554L.169S/abstract}{\apj}, 554, 169

\bibitem[{Sunyaev \& Truemper(1979)}]{1979Natur.279..506S}
Sunyaev, R.~A., \& Truemper, J. 1979, \href{https://ui.adsabs.harvard.edu/abs/1979Natur.279..506S/abstract}{Nature}, 279, 506

\bibitem[{Svensson \& Zdziarski(1994)}]{1994ApJ...436..599S}
Svensson, R., \& Zdziarski, A.~A. 1994, \href{https://ui.adsabs.harvard.edu/abs/1994ApJ...436..599S/abstract}{\apj}, 436, 599

\bibitem[{Svoboda {et~al.}(2012)Svoboda, Dovčiak, Goosmann, Jethwa, Karas, Miniutti, \& Guainazzi}]{2012A&A...545A.106S}
Svoboda, J., Dovčiak, M., Goosmann, R.~W., {et~al.} 2012, \href{https://ui.adsabs.harvard.edu/abs/2012A&A...545A.106S/abstract}{A\&A}, 545, 106

\bibitem[{Syunyaev {et~al.}(1994)Syunyaev, Borozdin, Aleksandrovich, Arefev, Kaniovskii, Efremov, Maisack, Reppin, \& Skinner}]{1994AstL...20..777S}
Syunyaev, R.~A., Borozdin, K.~N., Aleksandrovich, N.~L., {et~al.} 1994, \href{https://ui.adsabs.harvard.edu/abs/1994AstL...20..777S/abstract}{AstL}, 20, 890

\bibitem[{Tagger \& Pellat(1999)}]{1999A&A...349.1003T}
Tagger, M., \& Pellat, R. 1999, \href{https://ui.adsabs.harvard.edu/abs/1999A&A...349.1003T/abstract}{A\&A}, 349, 1003

\bibitem[{Tanaka \& Shibazaki(1996)}]{1996ARA&A..34..607T}
Tanaka, Y., \& Shibazaki, N. 1996, \href{https://ui.adsabs.harvard.edu/abs/1996ARA&A..34..607T/abstract}{ARA\&A}, 34, 607

\bibitem[{Tanaka {et~al.}(2007)Tanaka, Terashima, Torii, Ueda, Ushio, Watanabe, Yamauchi, \& Yaqoob}]{2007PASJ...59S.315M}
Tanaka, Y., Terashima, Y., Torii, K., {et~al.} 2007, \href{https://ui.adsabs.harvard.edu/abs/2007PASJ...59S.315M/abstract}{PASJ}, 59, 315

\bibitem[{Thorne \& Price(1975)}]{1975ApJ...195L.101T}
Thorne, K.~S., \& Price, R.~H. 1975, \href{https://ui.adsabs.harvard.edu/abs/1975ApJ...195L.101T/abstract}{\apj}, 195, 101

\bibitem[{Tripathi {et~al.}(2020)Tripathi, Liu, \& Bambi}]{2020MNRAS.498.3565T}
Tripathi, A., Liu, H., \& Bambi, C. 2020, \href{https://ui.adsabs.harvard.edu/abs/2020MNRAS.498.3565T/abstract}{\mnras}, 498, 3565

\bibitem[{Trudolyubov {et~al.}(2001)Trudolyubov, Borozdin, \& Priedhorsky}]{2001MNRAS.322..309T}
Trudolyubov, S.~P., Borozdin, K.~N., \& Priedhorsky, W.~C. 2001, \href{https://ui.adsabs.harvard.edu/abs/2001MNRAS.322..309T/abstract}{\mnras}, 322, 309

\bibitem[{van~den Eijnden {et~al.}(2017)van~den Eijnden, Ingram, Uttley, Motta, Belloni, \& Gardenier}]{2017MNRAS.464.2643V}
van~den Eijnden, J., Ingram, A., Uttley, P., {et~al.} 2017, \href{https://ui.adsabs.harvard.edu/abs/2017MNRAS.464.2643V/abstract}{\mnras}, 464, 2643

\bibitem[{van~der Klis(1989)}]{1989ASIC..262...27V}
van~der Klis, M. 1989, NATO ASI Series, Vol. 262, \href{https://ui.adsabs.harvard.edu/abs/1989ASIC..262...27V/abstract}{Fourier techniques in X-ray timing} (Kluwer Academic / Plenum Publishers)

\bibitem[{Vaughan \& Fabian(2002)}]{2004MNRAS.348.1415V}
Vaughan, S., \& Fabian, A.~C. 2002, \href{https://ui.adsabs.harvard.edu/abs/2004MNRAS.348.1415V/abstract}{\mnras}, 348, 1415

\bibitem[{Wagoner(1999)}]{1999PhR...311..259W}
Wagoner, R.~V. 1999, \href{https://ui.adsabs.harvard.edu/abs/1999PhR...311..259W/abstract}{Phys. Rep.}, 311, 259

\bibitem[{Walton {et~al.}(2017)Walton, Mooley, King, Tomsick, Miller, Dauser, García, Bachetti, Brightman, Fabian, Forster, Fürst, Gandhi, Grefenstette, Harrison, Madsen, Meier, Middleton, Natalucci, \& Rahoui}]{2017ApJ...839..110W}
Walton, D.~J., Mooley, K., King, A.~L., {et~al.} 2017, \href{https://ui.adsabs.harvard.edu/abs/2017ApJ...839..110W/abstract}{\apj}, 839, 110

\bibitem[{Wang {et~al.}(2021)Wang, Tang, Tuo, Epili, Zhang, Song, Lu, Qu, Zhang, Ge, Huang, Li, Bu, Cai, Cao, Chang, Chen, Chen, Chen, Chen, Chen, Cui, Du, Gao, Gao, Gu, Guan, Guo, Han, Huo, Jia, Jiang, Jin, Kong, Li, Li, Li, Li, Li, Li, Li, Li, Liang, Liao, Liu, Liu, Liu, Liu, Lu, Luo, Luo, Ma, Ma, Meng, Nang, Nie, Ou, Ren, Sai, Song, Sun, Tao, Wang, Wang, Wang, Wang, Wang, Wen, Wu, Wu, Wu, Xiao, Xiao, Xiong, Xu, Yang, Yang, Yang, Yang, Yi, Yin, You, Zhang, Zhang, Zhang, Zhang, Zhang, Zhang, Zhang, Zhang, Zhao, Zhao, Zheng, Zheng, \& Zhou}]{WANG20211}
Wang, W., Tang, Y., Tuo, Y., {et~al.} 2021, Journal of High Energy Astrophysics, 30, 1

\bibitem[{Weng {et~al.}(2018)Weng, Wang, Cai, Yuan, \& Gu}]{2018ApJ...865...19W}
Weng, S.-S., Wang, T.-T., Cai, J.-P., Yuan, Q.-R., \& Gu, W.-M. 2018, \href{https://ui.adsabs.harvard.edu/abs/2018ApJ...865...19W/abstract}{\apj}, 865, 19

\bibitem[{Wijnands {et~al.}(1999)Wijnands, Homan, \& van~der Klis}]{1999ApJ...526L..33W}
Wijnands, R., Homan, J., \& van~der Klis, M. 1999, \href{https://ui.adsabs.harvard.edu/abs/1999ApJ...526L..33W/abstract}{\apj}, 526, 33

\bibitem[{Wilkins \& Fabian(2011)}]{2011MNRAS.414.1269W}
Wilkins, D.~R., \& Fabian, A.~C. 2011, \href{https://ui.adsabs.harvard.edu/abs/2011MNRAS.414.1269W/abstract}{\mnras}, 414, 1269

\bibitem[{Wilkins \& Gallo(2013)}]{2015MNRAS.449..129W}
Wilkins, D.~R., \& Gallo, L.~C. 2013, \href{https://ui.adsabs.harvard.edu/abs/2015MNRAS.449..129W/abstract}{MNRAS}, 430, 1694

\bibitem[{Wilms {et~al.}(2000)Wilms, Allen, \& McCray}]{2000ApJ...542..914W}
Wilms, J., Allen, A., \& McCray, R. 2000, \href{https://ui.adsabs.harvard.edu/abs/2000ApJ...542..914W/abstract}{\apj}, 542, 914

\bibitem[{Yang {et~al.}(2022)Yang, Zhang, Huang, Bu, Zhang, Liu, Yu, Wang, Zhao, Tao, Qu, Zhang, Zhang, Song, Lu, Cao, Chen, Cai, Chang, \& Chen}]{2022ApJ...937...33Y}
Yang, Z.-x., Zhang, L., Huang, Y., {et~al.} 2022, \href{https://ui.adsabs.harvard.edu/abs/2022ApJ...937...33Y/abstract}{\apj}, 937, 33

\bibitem[{Zhang {et~al.}(2021)Zhang, Altamirano, Uttley, García, Méndez, Homan, Steiner, Alabarta, Buisson, Remillard, Gendreau, Arzoumanian, Markwardt, Strohmayer, Neilsen, \& Basak}]{2021MNRAS.505.3823Z}
Zhang, L., Altamirano, D., Uttley, P., {et~al.} 2021, \href{https://ui.adsabs.harvard.edu/abs/2021MNRAS.505.3823Z/abstract}{\mnras}, 505, 3823

\bibitem[{Zhang {et~al.}(2020)Zhang, Li, Lu, Song, Xu, Liu, Chen, Cao, Bu, Chang, Chen, Chen, Chen, Chen, Chen, Cui, Cui, Deng, Dong, \& Du}]{2020SCPMA..6349502Z}
Zhang, S.-N., Li, T., Lu, F., {et~al.} 2020, \href{https://ui.adsabs.harvard.edu/abs/2020SCPMA..6349502Z/abstract}{SCPMA}, 63, 249502

\bibitem[{Zhang {et~al.}(1995)Zhang, Jahoda, Swank, Morgan, \& Giles}]{1995ApJ...449..930Z}
Zhang, W., Jahoda, K., Swank, J.~H., Morgan, E.~H., \& Giles, A.~B. 1995, \href{https://ui.adsabs.harvard.edu/abs/1995ApJ...449..930Z/abstract}{\apj}, 449, 930

\bibitem[{Zhu \& Wang(2024)}]{2024ApJ...968..106Z}
Zhu, H., \& Wang, W. 2024, \href{https://ui.adsabs.harvard.edu/abs/2024ApJ...968..106Z/abstract}{\apj}, 968, 106

\bibitem[{Zoghbi {et~al.}(2010)Zoghbi, Fabian, Uttley, Miniutti, Gallo, Reynolds, Miller, \& Ponti}]{2010MNRAS.401.2419Z}
Zoghbi, A., Fabian, A.~C., Uttley, P., {et~al.} 2010, \href{https://ui.adsabs.harvard.edu/abs/2010MNRAS.401.2419Z/abstract}{\mnras}, 401, 2419

\end{thebibliography}
\bibliographystyle{aasjournal}

\end{document}